\documentclass[twocolumn]{aastex631}
\usepackage[T1]{fontenc}

\usepackage{gensymb}
\usepackage{CJK}
\usepackage{amsmath}

\begin{document}
\begin{CJK*}{UTF8}{gbsn}
\title{Swept by the Galactic Tide: How Galactic Tides and Stellar Encounters Drive the Evolution of Wide Binaries and Their Giant Planets}
\shorttitle{Swept by the Galactic Tide}
\shortauthors{Nagpal et al.}

\author[0000-0001-5909-4433]{Vighnesh Nagpal}
\altaffiliation{NSF Graduate Research Fellow}
\affiliation{Department of Astronomy \& Astrophysics, University of Chicago \\}

\author[0000-0003-0834-8645]{Tiger Lu (陆均)}
\altaffiliation{Flatiron Research Fellow}
\affiliation{Center for Computational Astrophysics, Flatiron Institute, 162 5th Avenue, New York, NY 10010, USA}

\author[0000-0002-7670-670X]{Malena Rice}
\affiliation{Department of Astronomy, Yale University, New Haven, CT 06511, USA}

\author[0000-0002-8532-827X]{Shaunak Modak}
\altaffiliation{Flatiron Research Fellow}
\affiliation{Center for Computational Astrophysics, Flatiron Institute, 162 5th Avenue, New York, NY 10010, USA}
\affiliation{Canadian Institute for Theoretical Astrophysics, University of Toronto, 60 St. George Street, Toronto, ON M5S 3H8, Canada}

\author[0000-0002-5861-5687]{Chris Hamilton}
\affiliation{Department of Astrophysical Sciences, Princeton University, 4 Ivy Lane, Princeton, NJ 08544, USA}

\author[0000-0003-0872-7098]{Adrian M. Price-Whelan}
\affiliation{Center for Computational Astrophysics, Flatiron Institute, 162 5th Avenue, New York, NY 10010, USA}

\author[0000-0002-4250-0957]{Diana Powell}
\affiliation{Department of Astronomy \& Astrophysics, University of Chicago \\}

\author[0000-0003-0012-9093]{Aida Behmard}
\altaffiliation{Flatiron Research Fellow}
\affiliation{Center for Computational Astrophysics, Flatiron Institute, 162 5th Avenue, New York, NY 10010, USA}






\begin{abstract}
Wide stellar binaries ($a_\mathrm{b} > 10^3$\,AU) are common throughout the Galaxy, and many host extrasolar planets. Comparing planets in wide binaries with those around single stars can reveal whether a distant stellar companion affects planet formation. Such a comparison cannot be made directly, however: unlike in single-star systems, the Galactic tide and passing field stars dramatically reshape wide binaries over Gyr timescales, influencing the orbital properties and stability of the planets they host. In this work, we present $N$-body simulations demonstrating how the Galactic tide and stellar encounters sculpt the orbits of super-Jupiters on 1--6.5\,AU orbits in wide binary systems on Sun-like Galactic orbits over 9\,Gyr. We predict systematically higher orbital eccentricities for planets in wide binaries compared to single-star systems, with the most significant differences in older systems and those with more massive binary companions. Planetary ejections and disruptions are also common, and we find that planet-hosting binaries with orbits nearly perpendicular to the Galactic midplane are preferentially depleted. In our most optimistic estimate, the maximum hot Jupiter production rate is enhanced by up to a factor of ${\sim}3$ relative to binaries without Galactic perturbations. A significant fraction of these binaries are also ionized (the stars become unbound), with a preference for ionization in lower-mass binaries, while the eccentricity distribution of survivors relaxes from mildly superthermal to nearly thermal. These results provide testable predictions for the population of giant planets in wide binaries soon to be revealed by Gaia DR4 and DR5.
\end{abstract}


\section{Introduction}
\label{sec:intro}

The architectures of planetary systems are commonly understood as the products of their formation and their subsequent dynamical evolution in isolation from the environments in which the systems are embedded. In reality, however, every planetary system orbits within the gravitational potential of the Galaxy, and over billions of years shares its neighborhood with a steady stream of passing stars.

The treatment of planetary systems as ``isolated'' is often a good approximation: planets themselves are typically bound far too tightly to their host stars to experience significant perturbations from the Galactic environment \citep[e.g.,][]{veras_evans_2013}. However, there are certain regimes in which the Galactic context drives qualitatively distinct evolutionary outcomes, such as at the fringes of planetary systems \citep[e.g.,][]{liveoak2026shaving}. In our Solar System, the icy bodies of the Oort cloud (tens of thousands of AU from the Sun) are continually nudged by the Galactic tide and passing stars, which periodically send them into the inner Solar System \citep{oort_1950, heisler_tremaine_86}---a process that is thought to be mirrored in extrasolar systems \citep{pham_polluting_white_dwarfs}.

Many planet-hosting stars also host a massive stellar companion, in some cases at separations comparable to the distance between the Sun and the Oort cloud. We define a ``wide binary" as a pair of stars in which both components are above the hydrogen-burning limit ($M \gtrsim 0.08\,M_\odot$; e.g., \citealt{chabrier_baraffe_2000}) and orbit at separations of $10^3\,\mathrm{AU} \lesssim a_\mathrm{b} \lesssim 10^5$\,AU. Such systems are common: ${\sim}5$--$10\%$ of FGK stars have a wide binary companion
\citep[e.g.,][]{lepine_bongiorno_2007, elbadry_catalog_21, gaia_edr3_nearby_stars_catalog}. At these separations, the binaries are so tenuously bound that perturbations by the Galactic environment critically shape their evolution. Over Gyr timescales the Galactic tide torques the binary's orbit, driving large-amplitude oscillations in its eccentricity and Galactic inclination \citep{heisler_tremaine_86, jiang_tremaine_10, hamilton_phase_mixed_22, modak_hamilton_tide_23}, while passing field stars stochastically perturb the orbit and can eventually disrupt the binary entirely \citep{jiang_tremaine_10, hamilton_modak_encounters_24}. 


There is strong evidence that closer binary companions ($a_\mathrm{b} < 10^3$\,AU) profoundly influence the planetary systems hosted by the primary star \citep[e.g.,][]{kraus_ruinous_2016}. Because the mechanisms by which binaries sculpt their planetary systems typically weaken steeply with increasing binary separation \citep[e.g.,][]{holman_wiegert, naoz_araa, gerbig2024aligning}, one might expect wider binaries to have comparatively little influence. However, \citet{kaib_2013} showed that this expectation breaks down for wide binaries once the Galactic environment is taken into account. The Galactic tide and passing stars may drive the binary's pericenter low enough to strongly perturb any planets in the system. \citet{kaib_2013} identified observational evidence supporting this hypothesis: planetary eccentricities in wide binary systems are systematically elevated compared to both single-star and close binary systems. More recently, \citet{grishin_wide_binary_hj_25} showed that the Galactic tide can enhance hot Jupiter formation in wide binaries, producing roughly twice as many hot Jupiters as in binaries without Galactic perturbations while ejecting a substantial fraction of giant planets entirely.

The upcoming Gaia \citep{{gaia2016}} data releases will soon allow these effects to be tested directly. Time-series astrometry from the fourth and fifth data releases (DR4 and DR5) of the Gaia mission is expected to reveal thousands of giant planets \citep{lammers_winn_gaia_26}, many of which will have stellar hosts with one or more wide stellar companions. The complementary Nancy Grace Roman Space Telescope's Galactic Bulge Time Domain Survey will, for the first time, provide a large sample of planetary systems well beyond the Solar neighborhood, spanning a wide range of Galactic environments out to the Galactic bulge \citep{penny_2019, wilson_roman_transiting}. Understanding how the Galaxy shapes planetary and binary systems will therefore become increasingly important for interpreting the planet populations that these surveys reveal.

Previous works have each captured only part of the relevant physics and parameter space. \citet{kaib_2013} used direct $N$-body simulations including both the Galactic tide and stellar encounters, but held the Galactic tide fixed at its local value and focused on Solar System-like configurations. \citet{grishin_wide_binary_hj_25} used a secular, orbit-averaged treatment of the vertical Galactic tide alone that did not account for stellar encounters, and considered planets at 5--200\,AU, largely beyond the region near the water-ice line where giant planets are most common \citep{fulton2021cls}. \citet{stegmann_24} evolved wide binaries under the influence of both the Galactic tide and stellar encounters without orbit averaging, finding that a significant fraction are driven to close approaches that produce stellar collisions, mergers, and gravitational-wave sources. However, they restricted their attention to massive stars and did not consider the impact of this binary evolution for planets in the systems.

In this work we combine the strengths of these previous approaches to study giant planets near the water-ice line: the population of exoplanets most accessible to Gaia DR4 and DR5. We follow planets drawn from simulated predictions of the Gaia planet yield, hosted by binaries drawn from observationally-motivated distributions, under a time-dependent Galactic tide computed along a realistic Galactic orbit and stellar encounters, resolving close pericenter passages with direct \textit{N}-body simulations where orbit-averaged approximations break down \citep{stegmann_24}. Finally, by comparing simulations with and without the Galactic tide and stellar encounters, we assess how each shapes the evolution of both the binaries and their planets.

The remainder of this paper is structured as follows. We describe our simulation setup in Section~\ref{methods} and discuss the evolution of wide binaries in the absence of a planet in Section~\ref{sec:binary_only}. We then outline the range of outcomes for the planet in Section~\ref{sec:binary_plus_planet}. We present population-level predictions for the eccentricities, survival, and orientations of planets in wide binaries detected by Gaia, as well as free-floating planet production, in Section~\ref{sec:planet_population}. Finally, we discuss caveats and implications of our results in Section~\ref{sec:discussion} and summarize our conclusions in Section~\ref{conclusion}.

\section{Methods}\label{methods}

\subsection{Simulation Setup}\label{sec:overview}

\begin{figure*}
    \centering
    \includegraphics[width=1.0\linewidth]{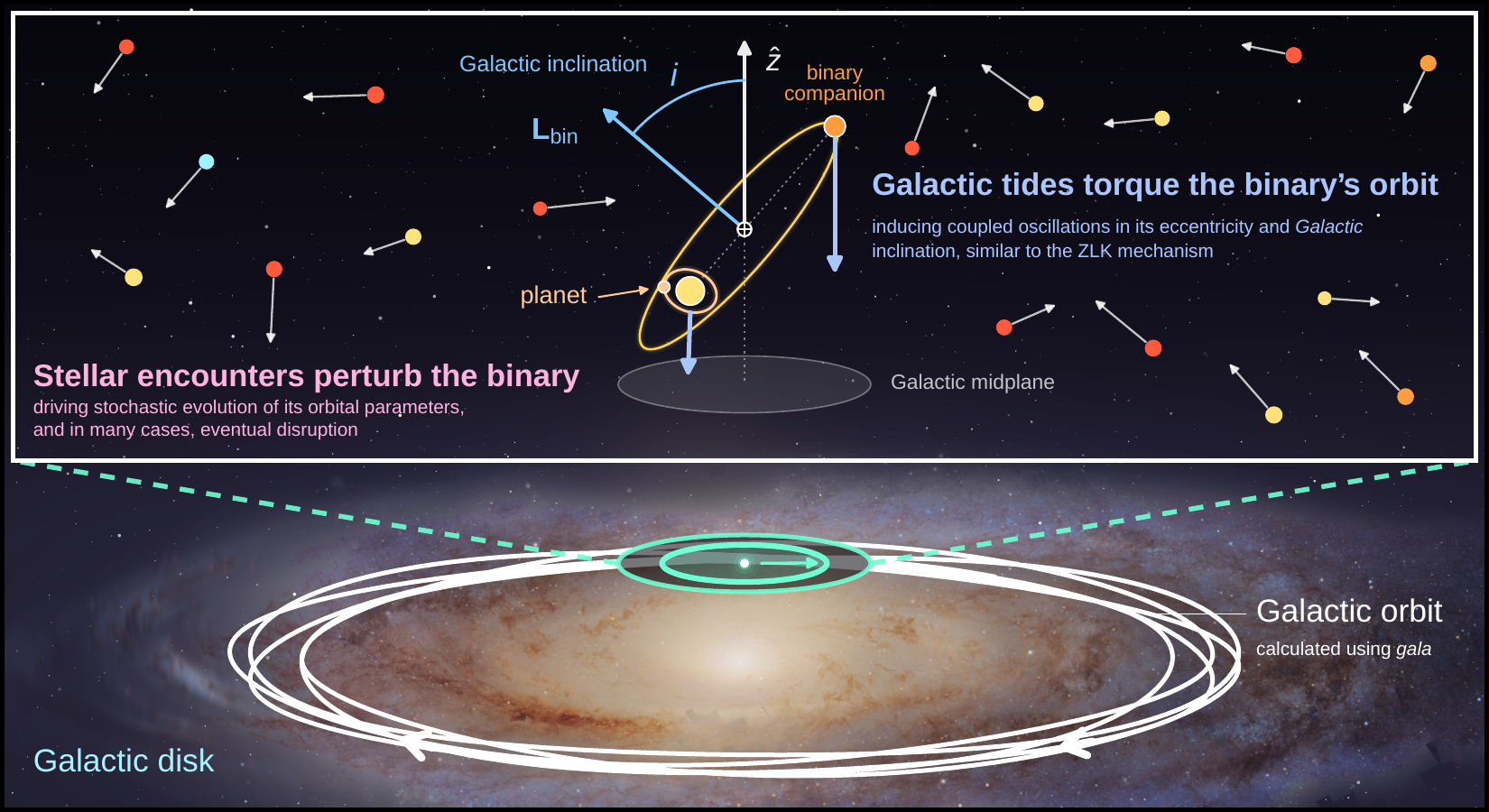}
    \caption{Schematic of a planet-hosting wide binary perturbed by the Galactic tide and stellar encounters. \textit{Bottom}: the system's orbit through the Galactic disk. \textit{Inset}: the Galactic tide torques the binary's orbit, driving coupled oscillations in its eccentricity and Galactic inclination relative to the Galactic disk's midplane, while passing field stars stochastically perturb and can eventually disrupt it. Not to scale. The Galactic orbit, which is calculated using a Milky Way potential (see Section~\ref{sec:tide}), is overlaid on an image of M31 (image credit: NASA, ESA, J. Dalcanton, B. F. Williams, L. C. Johnson, the PHAT team, and R. Gendler) for illustration.}
    \label{fig:schematic}
\end{figure*}

Figure~\ref{fig:schematic} depicts a schematic of the basic setup we consider in this work: a single super-Jupiter orbiting a star with a wide binary companion at an orbital separation between $10^3$ and $10^5$\,AU. At these separations, the companion star is only loosely bound to its primary, and its orbit may be significantly affected by the system's environment as it travels through the Galaxy.

We have run \textit{N}-body simulations of our setup with the \texttt{REBOUND}\footnote{\url{rebound.hanno-rein.de/}} package \citep{rein_rebound_11}. The setup of our simulations is as follows. In all simulations we include the leading-order general-relativistic apsidal precession induced by the primary star, using the \texttt{gr\_potential} implementation \citep{nobili_roxburgh_26} in \texttt{REBOUNDx} \citep{tamayo_reboundx_20}, a library for incorporating additional physics into \texttt{REBOUND}. To make predictions for the population of giant planets that Gaia DR4 and DR5 will be most sensitive to, we randomly sample the primary's stellar mass, planetary mass, and initial planetary orbital parameters from the Gaia DR5 mock catalog of \citet{lammers_winn_gaia_26}, which is built upon the planetary occurrence function from the California Legacy Survey \citep{rosenthal2021cls}. The host stars, drawn jointly with their planets from the same catalog entries, are predominantly K and G dwarfs: their masses span $0.1$--$2.0\,M_\odot$ with a median of $0.85\,M_\odot$ (16th--84th percentiles $0.59$--$1.08\,M_\odot$), and $89\%$ lie between $0.5$ and $1.5\,M_\odot$. We restrict our simulations to consider only planets with semimajor axes $> 1$\,AU. The resulting planets are predominantly massive giants near the water-ice line: their masses span $0.4$--$13\,M_{\rm J}$ with a median of $8.1\,M_{\rm J}$ (16th--84th percentiles $4.6$--$11.0\,M_{\rm J}$; 94\% exceed $3\,M_{\rm J}$), their semimajor axes lie between $1$ and $6.5$\,AU with a median of $3.2$\,AU (16th--84th percentiles $2.2$--$4.2$\,AU), and their eccentricities are drawn from the Beta distribution from \citealt{kipping_eccentricities_beta_distribution}. This region coincides with the expected peak of giant planet occurrence, near the location of the water-ice line for Sun-like stars \citep{ida2004deterministic, fernandes2019turnover, fulton2021cls}. 

The binary companion is initialized with semimajor axis drawn log-uniformly (i.e., according to \"Opik's law) between $10^3$--$10^5$\,AU, the separation range over which Galactic tides and stellar encounters are expected to be capable of driving significant orbital evolution \citep{hamilton_modak_encounters_24}. The companion mass is set as $M_2 = q\,M_1$, where $M_1$ is the mass of the primary and the mass ratio $q$ is drawn uniformly between $q_{\rm min} = \max(0.1,\, 0.08\,M_\odot/M_1)$ and 1, consistent with the roughly flat mass-ratio distribution observed for wide binaries \citep{moe_di_stefano_2017}. The lower limit is selected to keep the companion above the hydrogen-burning limit, while the upper limit ensures that the planet host is always the more massive star. The eccentricity of the wide binary companion is drawn from the separation-dependent power-law distribution inferred for the Gaia wide binary population by \citet{hwang_superthermal_22}, $p(e_\mathrm{b}) = (1+\alpha)\,e_\mathrm{b}^{\alpha}$ for $0 \le e_\mathrm{b} < 1$. Rather than adopting the power-law index $\alpha$ measured in a discrete separation bin, we evaluate their continuous fit to $\alpha$ as a function of separation at each system's initial binary semimajor axis, then draw $e_\mathrm{b}$ by inverse-transform sampling. Over our separation range, $\alpha$ increases from $1.19$ at $10^3$\,AU to $1.37$ at $10^5$\,AU, so that all of our binaries are drawn from superthermal distributions. The resulting eccentricities have a median of $0.73$, and $22\%$ exceed $0.9$. The binary's initial Galactic inclination (the angle between the binary's orbital plane and the Galactic midplane) and its initial mutual inclination relative to the planet are both isotropically distributed. The arguments of pericenter and longitudes of ascending node of both orbits are drawn uniformly from 0 to $2\pi$, and their initial orbital phases are drawn uniformly in mean anomaly.

All simulations use the \texttt{TRACE} integrator \citep{tiger_trace_24} in its wide-binary mode \citep{chambers_wb_integrator, tiger_trace_wb_26}. \texttt{TRACE} uses \texttt{REBOUND}'s implementation of the Wisdom--Holman method \citep{wisdom_holman_91, rein_tamayo_whfast_15} when possible, switching to a Bulirsch--Stoer (BS) integrator \citep{bs_integrator_press_numerical_recipes_07} as necessary to resolve close encounters and close pericenter passages. We use a global timestep equal to $0.06\times$ the planet's initial orbital period, reduced as needed so that it never exceeds 1/12 of the planet's instantaneous period, and integrate each system for 9\,Gyr. We set the BS tolerance parameter to $10^{-11}$ rather than the \texttt{TRACE} default of $10^{-8}$ in order to minimize secular drift. Every $0.8$\,Myr (our output cadence), we compute the osculating orbital elements of the planet with respect to the primary and of the companion with respect to the center of mass of the primary--planet pair (in Jacobi coordinates). If either eccentricity exceeds $1.05$, we consider the planet ejected or the binary ionized (that is, the two stars are no longer bound), respectively, and we terminate the simulation. The small margin above unity prevents a momentarily hyperbolic osculating orbit from ending the run prematurely. 

In our simulations, a planet is considered tidally disrupted if at any point during the integration\footnote{We implement this by giving the host star a collision radius equal to $a_{\rm R}$ and applying \texttt{REBOUND}'s line collision detection module, which detects passages within $a_{\rm R}$ even if they occur between timesteps.}
it passes within the Roche limit of its host star, which we take from the Roche-lobe approximation of \citet{eggleton_1983}
\begin{equation}
    a_\mathrm{R} = R_\mathrm{p} \, \frac{0.6\,(M_\mathrm{p}/M_1)^{2/3} + \ln\left[1 + (M_\mathrm{p}/M_1)^{1/3}\right]}{0.49\,(M_\mathrm{p}/M_1)^{2/3}},
    \label{eq:roche}
\end{equation} 
where $R_\mathrm{p}$ and $M_\mathrm{p}$ are the planet's radius and mass, respectively. For our planets this agrees to within ${\sim}3\%$ with the commonly used form $a_R = (R_\mathrm{p}/0.462)\,[M_\mathrm{p}/(M_1 + M_\mathrm{p})]^{-1/3}$ \citep{paczynski_araa_71, matsumura_tidal_evolution_2010}. Each planet is assigned a radius of 1\,$R_\mathrm{J}$, consistent with expectations from standard mass-radius relations \citep[e.g.,][]{chen2017radius}. In the case of a tidal disruption, we terminate the simulation.

We impose two influences from the Galactic environment. The first is the Galactic tide, which slowly torques the binary's orbit and drives secular oscillations in its eccentricity and Galactic inclination analogous to von Zeipel--Lidov--Kozai \citep[ZLK;][]{vonzeipel_1910, lidov_1962, kozai_1962, naoz_araa} oscillations. The second is encounters with passing field stars, which deliver kicks to the binary that cause its orbital elements to evolve stochastically and can eventually unbind the system. The specific implementation details of the Galactic tidal forcing and stellar encounters in our simulations are described in Sections~\ref{sec:tide} and \ref{sec:kicks}, respectively.

\subsection{Galactic Orbits and the Galactic Tide}\label{sec:tide}

The Galactic tide arises from the difference in Galactic acceleration across the system. At a position $\mathbf{r}$ relative to the binary's center of mass, this differential acceleration is, to leading order,
\begin{equation}
    \ddot{\mathbf{r}}^{\rm tide} = -\mathbf{H}\,\mathbf{r},
    \label{eq:tide}
\end{equation} where $\mathbf{H}$ is the Hessian of the Galactic potential $\Phi$ evaluated at the binary's center of mass, with components
\begin{equation}
    H_{ij} \equiv \frac{\partial^2 \Phi}{\partial x_i\, \partial x_j}
    \label{eq:hessian}
\end{equation}
where $x_i$ are the Cartesian coordinates of the system's position within the Galaxy.

\begin{figure}
    \centering
    \includegraphics[width=1.0\linewidth]{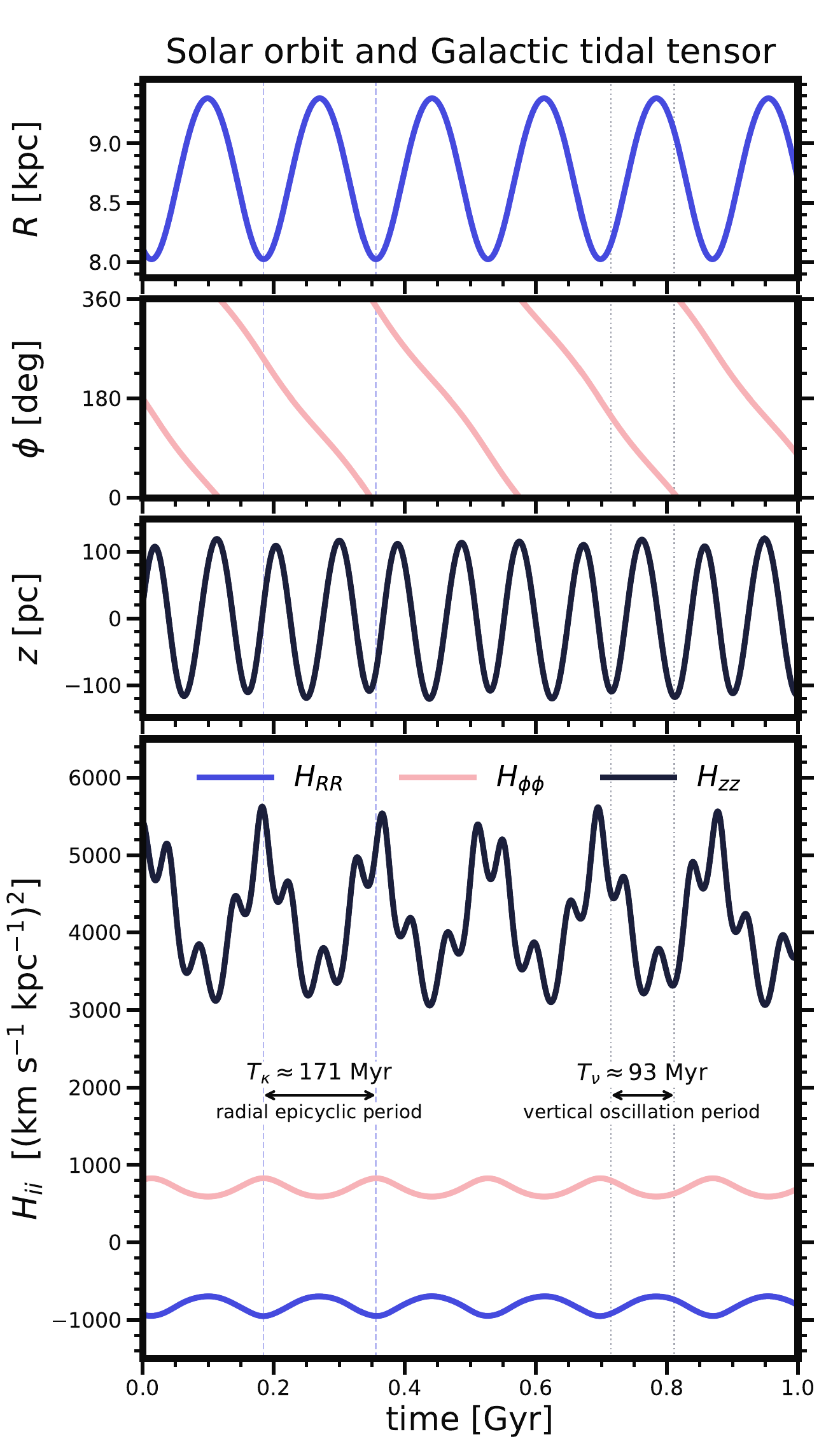}
    \caption{The Sun's orbit as integrated in \texttt{gala}'s \texttt{MilkyWayPotential} (top: $R$, $\phi$, $z$) and the diagonal components of the Galactic tidal tensor in the local $(R, \phi, z)$ frame (bottom), over 1\,Gyr from the present day. All three components vary mainly with the radial epicyclic period $T_\kappa$ (dashed lines, at radial pericenters). $H_{zz}$ also peaks at every midplane crossing, twice per vertical period $T_\nu$ (dotted lines, at minima of $z$).}
    \label{fig:galactic_orbit_and_tensor}
\end{figure}

We precompute the Galactic orbit of each binary's center of mass over 9\,Gyr (the approximate age of the Galactic thin disk; e.g., \citealt{nordstrom_2004,haywood_2013,xiang_rix_2022}) using the \texttt{gala} package \citep{gala_17} and its built-in \texttt{MilkyWayPotential} model \citep{bovy_galpy_15}. Every simulated system in this work follows the Sun's orbit through this potential, which we obtain by integrating forward from the Sun's present-day position and velocity ($R_0 = 8.122$\,kpc, $z_\odot = 20.8$\,pc, and $\mathbf{v}_\odot = [12.9, 245.6, 7.78]$\,km\,s$^{-1}$; e.g., \citealt{GRAVITY_2018_R0, drimmel_poggio_solar_velocity, bennett_bovy_2019}). The resulting orbit, shown in the top three panels of Figure~\ref{fig:galactic_orbit_and_tensor} over the first Gyr, spans Galactocentric radii of $8.0$--$9.4$\,kpc and reaches a maximum height of ${\sim}120$\,pc above the midplane. Although the potential in our setup is itself static, the Galactic tide experienced by each binary is time-dependent, since the system samples a range of Galactocentric radii and heights above the midplane as it moves along its Galactic orbit.

The bottom panel of Figure~\ref{fig:galactic_orbit_and_tensor} shows the diagonal components of $H_{ij}$ along the Sun's orbit in this local $(\hat{R}, \hat{\phi}, \hat{z})$ basis. In the Solar neighborhood, the vertical component dominates, with $H_{zz}$ several times larger in magnitude than $H_{RR}$ or $H_{\phi\phi}$, as expected for a thin Galactic disk in which vertical density gradients are much steeper than planar ones. All three components vary mainly over the radial epicyclic period, $T_\kappa \approx 171$\,Myr. $H_{zz}$ additionally peaks at every midplane crossing: that is, twice per vertical oscillation period $T_\nu \approx 93$\,Myr. Both timescales are much shorter than the secular timescale over which the tide reshapes the binary (Section~\ref{sec:binary_only}), so the binary effectively responds to the orbit-averaged tide, modulated by these oscillations.

Since we are interested in the orbits of the planet and its stellar companion, which span scales of AU to ${\sim}10^5$\,AU, we perform our \textit{N}-body integrations in a frame centered on the binary's center of mass, rather than in a Galactocentric frame where these orbits would be dwarfed by the system's ${\sim}8$\,kpc distance from the Galactic Center. We orient this frame so that its axes stay aligned with the local $(\hat{R}, \hat{\phi}, \hat{z})$ directions, the same basis in which we express $\mathbf{H}$. Because this frame both follows the binary along its Galactic orbit and rotates with it, it is non-inertial: it rotates with angular velocity $\boldsymbol{\Omega}_g = \dot{\phi}\,\hat{z}$, where $\dot{\phi}$ is the instantaneous Galactocentric angular speed of the system, calculated from the \texttt{gala} output. We account for this by adding the corresponding fictitious forces, so that the equation of motion for each body becomes
\begin{equation}
    \ddot{\mathbf{r}} = \mathbf{a}_{\rm int}
    - \mathbf{H}\,\mathbf{r}
    - 2\,\boldsymbol{\Omega}_g \times \dot{\mathbf{r}}
    - \boldsymbol{\Omega}_g \times \left(\boldsymbol{\Omega}_g \times \mathbf{r}\right)
    - \dot{\boldsymbol{\Omega}}_g \times \mathbf{r},
    \label{eq:eom}
\end{equation} where $\mathbf{a}_{\rm int}$ is the gravitational acceleration from the other bodies in the system. The last three terms are the Coriolis, centrifugal, and Euler forces, respectively. The Euler term is nonzero because the Sun's orbit is not circular, so $\dot{\phi}$ varies along the orbit.

We tabulate $\mathbf{H}$ and $\Omega_g$ every 0.05\,Myr along the precomputed Galactic orbit and, at every timestep, linearly interpolate them to the simulation time. We add the resulting Galactic tidal and fictitious accelerations (Equation~\ref{eq:eom}) to each body as an additional force applied every timestep using \texttt{REBOUNDx}.

\subsection{Stellar Encounters}\label{sec:kicks}
Ideally, we would directly model stellar encounters by introducing a passing star into our \textit{N}-body simulations. Unfortunately, the structure of the \texttt{TRACE} integrator in wide-binary mode does not support direct integrations that include both a wide binary companion and a passing star. 

We hence treat stellar encounters following \citet{binney_tremaine_textbook} and \citet{jiang_tremaine_10}, modeling their cumulative effect as a diffusive random walk in the velocities of the binary components rather than simulating individual passing stars. This treatment relies on three assumptions, which \citet{hamilton_modak_encounters_24} justify in detail for wide binaries in the Solar neighborhood. First, we assume that encounters are \textit{impulsive}. Since field stars pass the binary at a relative velocity dispersion of $\sigma \approx 40$\,km\,s$^{-1}$ \citep[e.g.,][]{dehnen_binney_98}, much faster than a binary's typical orbital velocity ($\lesssim 1$\,km\,s$^{-1}$), each encounter is brief compared to the binary's orbital period, and thus the stars change in velocity but not substantially in position. Second, we assume that encounters are \textit{weak}, i.e.\ that the velocity change from any single encounter is small compared to the binary's orbital velocity. Encounters strong enough to violate this assumption are rare and are only important for the very widest binaries. Third, we assume that the binary experiences many \textit{penetrative} encounters over its lifetime, i.e.\ encounters with impact parameters $b \lesssim a_\mathrm{b}$ \citep{hamilton_modak_encounters_24}. This assumption is what makes the binary's evolution diffusive, since it is then driven by many uncorrelated kicks rather than being dominated by the single closest encounter it experiences. We refer the interested reader to \citet{caputo_tomaselli_hamilton_2026} for a detailed treatment of binary evolution under stellar encounters across all of these regimes. Hereafter, we use the terms stellar encounter and ``kick" interchangeably.

Explicitly, every $\Delta t = 0.8$\,Myr (our output cadence) we apply an independent random velocity kick to each of the two components of the wide binary: the companion with mass $M_2$, and the inner star--planet pair with total mass $M_1 + M_\mathrm{p}$. The primary star and planet receive identical kicks, so encounters move the pair as a whole but do not directly perturb the planet's orbit about its host. For component $i$, with mass $M_i$ and velocity $\textbf{v}_i$ (magnitude $v_i \equiv |\mathbf{v}_i|$) relative to the binary's center of mass, we draw each Cartesian component of the kick independently from a Gaussian distribution whose mean represents a systemic velocity change due to dynamical friction, and the variance corresponds to random scatter from stellar encounters. Both are described by diffusion coefficients \citep[see][]{binney_tremaine_textbook}. In an isotropic Maxwellian field of perturbers with velocity dispersion $\sigma$, the diffusion coefficients are most naturally expressed in components parallel $(\parallel)$ and perpendicular $(\perp)$ to $\textbf{v}_i$. In the limit where $v_i \ll \sigma$, Equation~31 of \citet{jiang_tremaine_10} gives

\begin{equation}
    D[\Delta v_\parallel] = -\frac{4\sqrt{2\pi}\,G^2 (M_i \rho_1 + \rho_2) \ln\Lambda}{3\sigma^3}\, v_i,
    \label{eq:dfric}
\end{equation}
\begin{equation}
    D[(\Delta v_\parallel)^2] = \frac{8\sqrt{2\pi}\,G^2 \rho_2 \ln\Lambda}{3\sigma},
    \label{eq:dpar}
\end{equation}
\begin{equation}
    D[(\Delta v_\perp)^2] = 2\,D[(\Delta v_\parallel)^2],
    \label{eq:dperp}
\end{equation}
where $\rho_1$ and $\rho_2$ are moments of the perturber mass function---in our case the mass function from \citet{kroupa_mass_function_1993}---and $\ln\Lambda = \ln (b_\mathrm{max}/b_\mathrm{min})$ is the Coulomb logarithm, which accounts for the range of impact parameters $b$ contributing to the diffusion. Following \citet{jiang_tremaine_10}, we take $b_{\rm min} = G(M_i + M_\mathrm{pert})/\sigma^2 \approx 1$\,AU with a perturber mass $M_\mathrm{pert} = 1\,M_\odot$, and set $b_{\rm max}$ to half the instantaneous binary separation, recomputed at every kick epoch. This gives $\ln\Lambda \approx 6$--$11$ across our grid.

Physically, Equation~\ref{eq:dfric} represents dynamical friction, directed in the $-\hat{\textbf{v}}_i$ direction, opposite the body's velocity. Because $D[(\Delta v_\perp)^2]$ sums the contributions of the two directions perpendicular to $\textbf{v}_i$, Equation~\ref{eq:dperp} implies that each perpendicular direction receives the same mean-square velocity change, $D[(\Delta v_\parallel)^2]$, as the parallel direction. The random part of each kick is therefore isotropic, with equal variance in every direction. We draw each Cartesian component of the kick independently from a Gaussian whose mean is the corresponding component of $D[\Delta v_\parallel]\,\Delta t\,\hat{\mathbf{v}}_i$ and whose variance is $D[(\Delta v_\parallel)^2]\,\Delta t$.

We then subtract the mass-weighted mean kick from each body. Since all external forces in Equation~\ref{eq:eom} are linear in position and velocity, this changes no relative orbit, but it prevents the center of mass from random-walking far from the origin and accumulating roundoff error in the planet's orbit. Because the kick moments scale linearly with $\Delta t$, the accumulated kick is independent of our choice of $\Delta t$, provided $\Delta t$ is shorter than the characteristic evolution timescale of the binary orbit but long enough that treating the cumulative effect of weak kicks as a Gaussian diffusion process is justified.

\begin{figure*}
    \centering
    \includegraphics[width=1.0\linewidth]{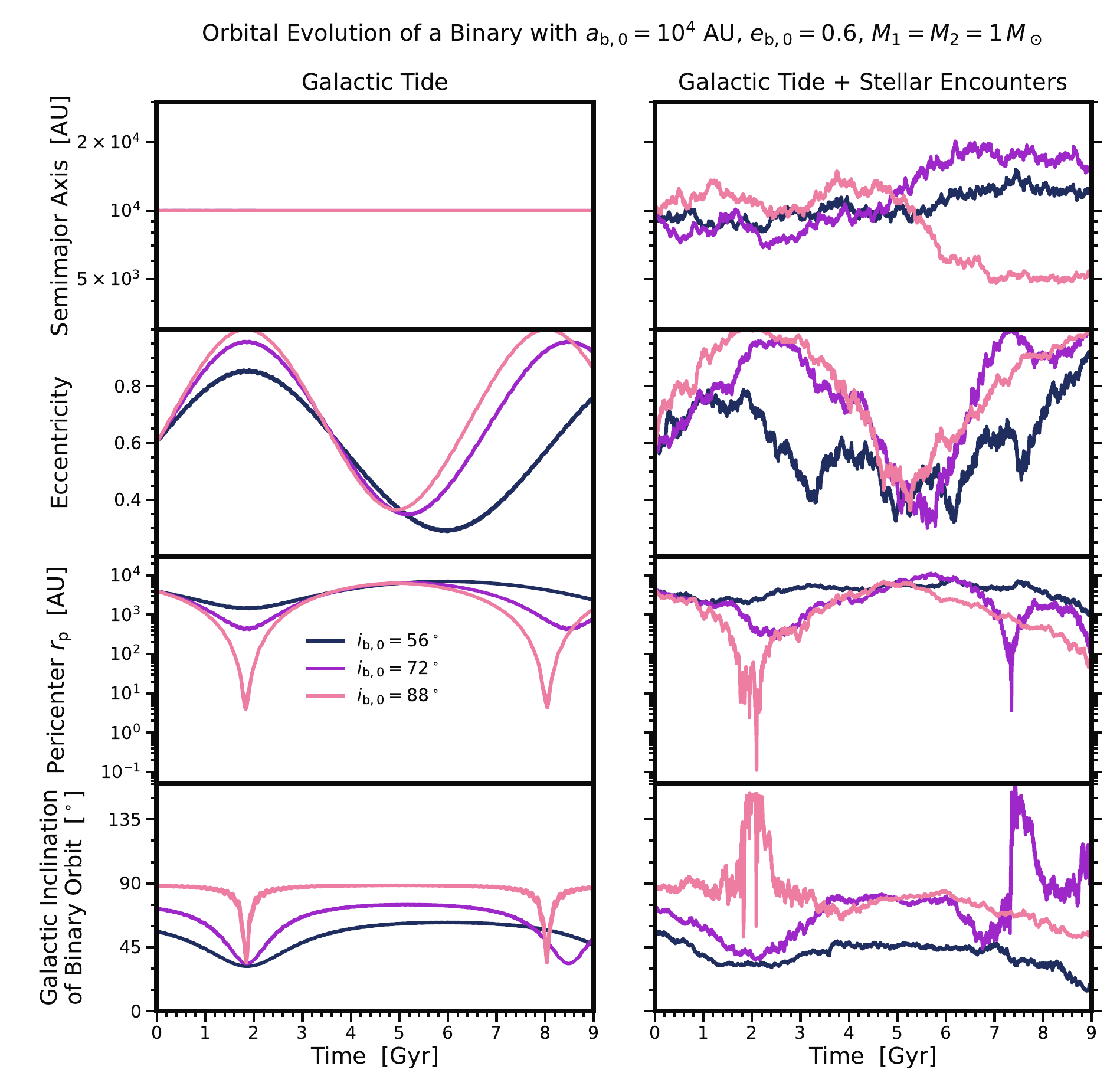}
    \caption{Orbital evolution of a Solar-twin binary ($a_{\mathrm{b},0} = 10^4$\,AU, $e_{\mathrm{b},0} = 0.6$, $M_1 = M_2 = 1\,M_\odot$) over 9\,Gyr for three initial Galactic inclinations, with all other initial parameters held fixed. Rows show the binary's semimajor axis, eccentricity, pericentre distance, and Galactic inclination. \textit{Left:} Under the Galactic tide alone, each binary undergoes smooth, periodic oscillations, with higher initial inclinations reaching higher eccentricities and closer pericentres. The semimajor axis remains fixed, leading to the overlapping lines in the top left panel. \textit{Right:} Stellar encounters overlay stochastic evolution on these oscillations, allowing binaries to reach closer pericentres than the tide alone permits.}
    \label{fig:galactic_tide_vs_encounters_demonstration}
\end{figure*}

\subsection{Simulation Grids}\label{sec:ics}
We ran three grids of simulations, each encompassing a set of 2500 identical initial conditions for planet-hosting wide binaries. The three grids include the following physics, respectively:
\begin{enumerate}
    \item No Galactic environmental effects,
    \item The Galactic tide alone,
    \item Both the Galactic tide and stellar encounters.
\end{enumerate}  
We additionally run the two grids with Galactic perturbations with the planet removed to provide a planet-free baseline for the evolution of the binaries. Although the planet might naively be expected to have little effect on the binary, a super-Jupiter at a few AU is typically much more tightly bound to its host than the wide companion is. In some cases, therefore, an exchange of energy between the two during a close approach during a pericenter dip can significantly alter or even unbind the binary's orbit (Section~\ref{sec:planets_ionize}).

We therefore begin, in Section~\ref{sec:binary_only}, by discussing the evolution of wide binaries in the absence of a planet, before turning to how planets respond to this evolution in Section~\ref{sec:binary_plus_planet}.

\section{Binary Evolution}\label{sec:binary_only}

In this section, we outline how wide binary orbits evolve under the influence of the Galactic tide and stellar encounters in the absence of a planet, using the binary-only simulations described in Section~\ref{sec:ics}. Since the planet is affected primarily through the evolution of its host binary, rather than through direct perturbation from Galactic-level effects, understanding how the binary orbit evolves provides crucial context for interpreting the planet's dynamical response that is discussed in Section~\ref{sec:binary_plus_planet}.

We first provide theoretical background for how a binary orbit is sculpted by the Galactic tide, both in isolation (Section~\ref{sec:binary_tide}) and in conjunction with stellar encounters (Section~\ref{sec:binary_kicks}). Then, in Section~\ref{sec:binary_pop} we report the population-level consequences of both these effects from our \textit{N}-body simulations. Unless otherwise noted, the population-level results in this section are drawn from the binary-only grid with both the Galactic tide and stellar encounters included.

\subsection{The Galactic Tide}\label{sec:binary_tide}
 
The Galactic tide drives oscillations in the eccentricity and Galactic inclination of the binary's orbit over multi-Gyr timescales. While our \textit{N}-body simulations make no orbit-averaging assumptions, the secular theory of how a binary evolves under the influence of the Galactic tide offers valuable physical intuition for interpreting our results, so we briefly summarize it here.

In particular, we can gain insight into these oscillations by averaging over both the binary's Keplerian motion (typical periods of ${\sim}1\,$Myr) and its Galactic orbit (typical periods of ${\sim} 100\,$Myr), retaining only the linear tidal effect of the Galactic potential. In this doubly-averaged, quadrupole limit, the binary's semimajor axis $a_\mathrm{b}$ remains constant. The component of the binary's dimensionless angular momentum vector perpendicular to the Galactic midplane $j_z$ is also conserved as
\begin{equation}
    j_z \equiv \sqrt{1 - e_\mathrm{b}^2}\cos i_\mathrm{b} = \mathrm{\,constant,}
    \label{eq:jz}
\end{equation}
where $e_\mathrm{b}$ is the binary's orbital eccentricity and $i_\mathrm{b}$ its Galactic inclination (e.g., \citealt{heisler_tremaine_86}). The binary's eccentricity and Galactic inclination are thus traded off against each other, analogous to the well-studied ZLK mechanism \citep{hamilton2019secular}.

Following \cite{modak_hamilton_tide_23}, the period of these oscillations $t_{\mathrm{sec}}$ can be approximated as
\begin{equation}
\begin{split}
    t_{\rm sec} \approx & 1\,{\rm Gyr}
    \left(\frac{\rho_0}{0.2\,M_\odot\,{\rm pc}^{-3}}\right)^{-1}\\
    & \times\left(\frac{M_\mathrm{bin}}{M_\odot}\right)^{1/2}
    \left(\frac{a_\mathrm{b}}{10^4\,{\rm AU}}\right)^{-3/2},
    \end{split}
    \label{eq:tsec}
\end{equation} where $\rho_{0}$ is the total local mass density in the Galactic midplane, including stars, stellar remnants, gas, and dark matter (often called the dynamical density), and $M_\mathrm{bin}$ is the total mass of the binary.

The conservation of $j_z$ provides useful intuition for which binaries the tide can drive to close pericenter. Since $\sqrt{1 - e_\mathrm{b}^2} \geq |j_z|$, a binary can reach an eccentricity no higher than $e_{\rm max} = \sqrt{1 - j_z^2}$, corresponding to a minimum binary pericenter:
\begin{equation}
    a_\mathrm{b}(1-e_\mathrm{b}) \gtrsim \frac{a_\mathrm{b}\, j_z^2}{2}
    = \frac{a_\mathrm{b}}{2}\left(1 - e_{\mathrm{b}, 0}^2\right)\cos^2 i_{\mathrm{b}, 0}
    \label{eq:qmin}
\end{equation}
for small $j_z$, where $e_{\mathrm{b}, 0}$ and $i_{\mathrm{b}, 0}$ are the binary's initial eccentricity and Galactic inclination. Reaching close pericenter passages therefore requires a low $j_z$, which can arise either from an orbit nearly perpendicular to the Galactic midplane or from an already-high eccentricity. In practice, the Galactic inclination is the more powerful lever, since at high Galactic inclination $\cos^2 i_{\mathrm{b}, 0}$ can approach zero for any initial eccentricity.

While this picture provides useful physical intuition, it is nonetheless a first-order idealization. Equation~\ref{eq:qmin} assumes a purely vertical, time-averaged Galactic tide, whereas the Galactic tide in our simulations varies along the system's Galactic orbit (Figure~\ref{fig:galactic_orbit_and_tensor}) and includes smaller planar components, both of which cause $j_z$ to fluctuate rather than remain strictly constant. Our simulations also capture effects beyond the orbit-averaged, leading-order treatment described above. Most importantly, stellar encounters disrupt the conservation of $j_z$ altogether (Section~\ref{sec:binary_kicks}). Equation~\ref{eq:qmin} should therefore be viewed as a guide to which binaries are most susceptible to the tide, rather than as a strict limit on their evolution.

Figure~\ref{fig:galactic_tide_vs_encounters_demonstration} illustrates the essential physics, using a Solar-twin binary with $a_\mathrm{b} = 10^4$\,AU and $e_{\mathrm{b},0} = 0.6$ at three initial Galactic inclinations, with all other initial orbital parameters held fixed. In the tide-only case (left), each binary undergoes smooth oscillations with a period of ${\sim}6$\,Gyr, of the same order as $t_{\rm sec}$ (Equation~\ref{eq:tsec}). As expected, binaries with higher initial Galactic inclinations reach higher eccentricities and substantially lower pericenters, with the eccentricity rising as the Galactic inclination falls, following Equation~\ref{eq:jz}. A small change in a binary's initial Galactic inclination can thus change its pericenter approach by orders of magnitude, with dramatic consequences for the planet (Sections~\ref{sec:binary_plus_planet} and \ref{sec:planet_population}).

\subsection{Stellar Encounters}\label{sec:binary_kicks}

Wide binaries do not evolve under the Galactic tide alone: stellar encounters also play an important role. Each kick changes the binary's orbital angular momentum, breaking the conservation of $j_z$ that constrains purely tide-driven evolution. Consequently, the binary's orbit random-walks on top of the broad trend set by the Galactic tide. The right column of Figure~\ref{fig:galactic_tide_vs_encounters_demonstration} shows this for the same three binaries as described in Section \ref{sec:binary_tide}: the kicks add significant stochasticity on top of the still-visible tide-driven oscillations. Importantly, since encounters can lower $j_z$, the binaries' pericenter can fall below the minimum dictated by Equation~\ref{eq:qmin}. 

In addition, encounters are most effective at reorienting the binary's orbit when it is highly eccentric. A kick of a given size changes the angular momentum by roughly the same amount regardless of eccentricity, but, for a fixed semimajor axis, the angular momentum itself becomes small as $e_\mathrm{b} \to 1$. Because the tide-driven pericenter minima coincide with the eccentricity maxima, even modest kicks during these phases can substantially tilt the orbit, producing the abrupt jumps in Galactic inclination seen in Figure~\ref{fig:galactic_tide_vs_encounters_demonstration}. Encounters can therefore move a binary into, or out of, the range of Galactic inclinations for which the tide drives it to high eccentricity, giving binaries of any initial inclination a route to close pericenter passages. This weakens the inclination dependence that tide-driven evolution alone would produce (see Section~\ref{sec:planet_population}).

Over Gyr timescales, stellar encounters gradually pump energy into the binary population such that binaries on average widen until they are ionized (in our simulations, when the companion's eccentricity exceeds $1.05$; Section~\ref{sec:overview}). The Galactic tide is often responsible for the final unbinding, which quickly strips binaries once their separations approach the Jacobi radius (${\sim}10^{5}$\,AU). This hastens ionization but does not significantly change the final ionization fraction, which is set by the encounters. The typical timescale for encounter-driven ionization is given by \cite{binney_tremaine_textbook, jiang_tremaine_10}
\begin{equation}
    t_{\rm dis} \approx \frac{0.02\,\sigma\, M_\mathrm{bin}}{G\rho_2\, a_\mathrm{b} \ln\Lambda},
    \label{eq:tdis}
\end{equation}
where $\sigma$, $\rho_2$, and $\ln\Lambda$ are as defined in Section~\ref{sec:kicks}. Wider and lower-mass binaries are less tightly bound and thus ionize more quickly. For the field-star parameters we adopt, whose $\rho_2$ is consistent with the local stellar census \citep{mckee_parravano_hollenbach_2015}, this evaluates to
\begin{equation}
\begin{split}
    t_{\rm dis} \approx & 13\,{\rm Gyr}
    \left(\frac{\ln\Lambda}{10}\right)^{-1}
    \left(\frac{\rho_2}{0.029\,M_\odot^2\,{\rm pc}^{-3}}\right)^{-1}
    \left(\frac{\sigma}{40\,{\rm km\,s^{-1}}}\right)\\
    & \times\left(\frac{M_\mathrm{bin}}{M_\odot}\right)
    \left(\frac{a_\mathrm{b}}{10^4\,{\rm AU}}\right)^{-1}.
    \end{split}
    \label{eq:tdis_num}
\end{equation} We examine how these dependencies shape the population of wide binaries in Section~\ref{sec:binary_pop}.

\begin{figure}
    \centering
    \includegraphics[width=1.0\linewidth]{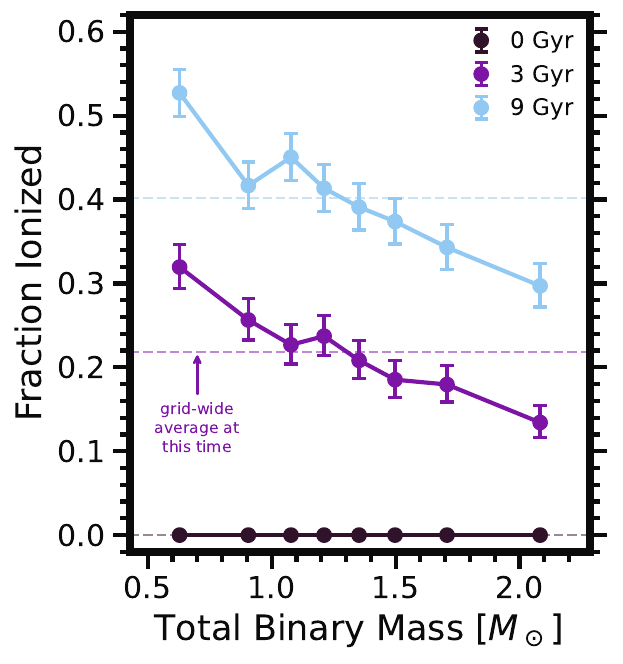}
    \caption{Fraction of wide binaries ionized by 0, 3 and 9 Gyr (under the Galactic tide + stellar encounters) as a function of total binary mass. Dashed lines mark the grid-wide average: the fraction of all 2500 binaries, regardless of mass, ionized by that time.} 
    \label{fig:binary_disruption_population_level}
\end{figure}

\subsection{Population-level Simulation Results}\label{sec:binary_pop}
Binary ionizations are a common outcome of our simulations. Approximately 22\% of our binaries are ionized by 3\,Gyr and 40\% by 9\,Gyr. We find that binaries with a lower total mass are preferentially ionized (Figure~\ref{fig:binary_disruption_population_level}), with ${\sim}33\%$ of binaries with $M_\mathrm{bin}\approx0.6\,M_\odot$ (which are predominantly M-M binaries) ionized by 3\,Gyr, in contrast to only ${\sim}13\%$ of binaries with $M_\mathrm{bin}\approx 2\,M_\odot$ at the same epoch. This trend persists to late times, with the ionized fractions climbing to 53\% and 30\% respectively by 9\,Gyr.

Galactic perturbations preferentially remove lower-mass wide binaries, systematically shifting the surviving population toward higher masses over time (Figure~\ref{fig:binary_disruption_population_level}). The finding that lower-mass wide binaries are more easily ionized is expected from classic calculations of binary ionization by passing stars \citep[e.g.,][]{bahcall_hut_tremaine_85, weinberg_1987, jiang_tremaine_10}, and dynamical ionization has long been discussed as a possible explanation for the observed paucity of wide, low-mass binaries \citep[e.g.,][]{burgasser_vlms_multiplicity}. If wide binaries of all masses were born with the same occurrence rate, the ionizations in our simulations alone would leave ${\sim}1.3$--$1.5$ times as many surviving wide binaries with $M_\mathrm{bin} \approx 2\,M_\odot$ as with $M_\mathrm{bin} \approx 0.6\,M_\odot$ after several Gyr. For comparison, the observed wide binary fraction is ${\sim}3$--$3.5$ times higher for F and G stars than for M dwarfs \citep{gaia_edr3_nearby_stars_catalog}. Dynamical ionization may therefore account for a modest but non-negligible part of this trend, with the remainder likely imprinted at formation, consistent with evidence that the mass dependence of wide binary occurrence is already present in young populations \citep[e.g.,][]{kraus_2009_widebinary}.

\begin{figure}
    \centering
    \includegraphics[width=1.0\linewidth]{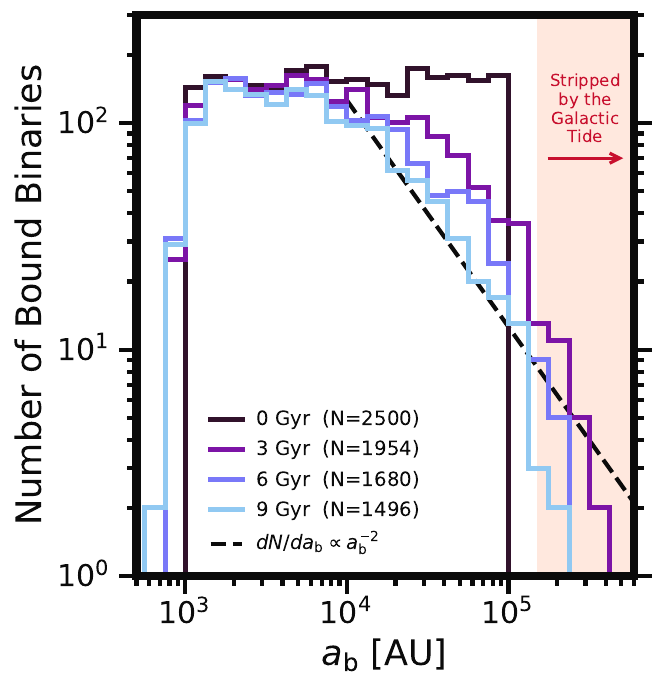}
    \caption{The evolution of the binary semimajor axis distribution from 0 to 9 Gyr starting with a log-uniform distribution. Only bound binaries are counted, so the number of binaries $N$ decreases over time as binaries are ionized by stellar encounters. For reference, we overplot a power-law distribution $dN/da_\mathrm{b} \propto a_\mathrm{b}^{-2}$ as a black dashed line. }
    \label{fig:binary_sma_evolution}
\end{figure}

Our binary semimajor axis distribution also evolves. We show the number of bound binaries at each semimajor axis at the same timestamps in Figure~\ref{fig:binary_sma_evolution}. Bound wide binaries are gradually heated by stellar encounters, random-walking to larger separations where they are more easily ionized. Since this process operates more efficiently at larger $a_\mathrm{b}$, our initially log-uniform distribution of semimajor axes progressively develops a tail following the distribution $dN/da_{\rm b}\propto a_\mathrm{b}^{-2}$ at large semimajor axes\footnote{This power-law distribution is known to be an approximate steady-state distribution of the Fokker-Planck equation governing semimajor axis evolution under weak, impulsive stellar encounters (see, e.g., Equation 147 of \citealt{caputo_tomaselli_hamilton_2026})}---reminiscent of the observed field distribution \citep{tian_elbadry_separation_2019,elbadry_catalog_21}.

\begin{figure}
    \centering
    \includegraphics[width=1.0\linewidth]{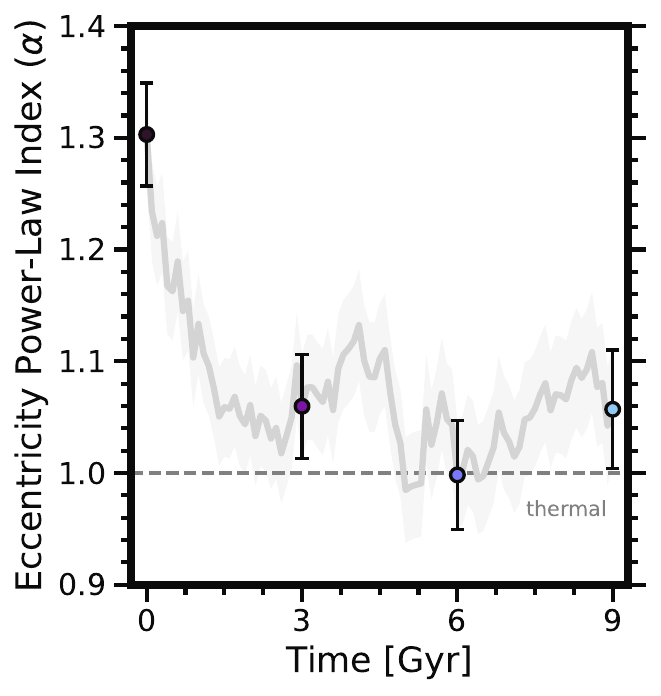}
    \caption{The best-fit power-law index $\alpha$ of the eccentricity distribution, $f(e_\mathrm{b}) \propto e_\mathrm{b}^{\alpha}$, of the wide binaries in our binary-only grid with both the Galactic tide and stellar encounters, as a function of time. At each time, $\alpha$ is the maximum-likelihood fit to the binaries still bound at that time. The gray line shows this fit every 0.1\,Gyr, with the shaded band showing its $1\sigma$ statistical uncertainty. The colored points mark the fits at 0, 3, 6, and 9\,Gyr, with $1\sigma$ error bars. The dashed line indicates a thermal distribution ($\alpha = 1$). The tide and stellar encounters thermalize the distribution from its superthermal initialization within ${\sim}1.5$\,Gyr.} 
    \label{fig:binary_ecc_dist_evolution}
\end{figure}

Unlike the semimajor-axis distribution, which continuously loses its widest binaries as the integrations progress, the eccentricity distribution evolves rapidly over the first few Gyr before settling to a near-thermal distribution (Figure~\ref{fig:binary_ecc_dist_evolution}). Every 0.1\,Gyr, we calculate the osculating eccentricities of all binaries that remain bound and fit the eccentricities with a power law, $dN/de_\mathrm{b} = (\alpha+1)\,e_\mathrm{b}^\alpha$ on $0 \le e_\mathrm{b} < 1$, by unbinned maximum likelihood. This fits the simulated eccentricities directly rather than the projected separation--velocity angles used by \citet{hwang_superthermal_22}, so the two estimators may differ slightly. We find that $\alpha$ relaxes from the mildly superthermal distribution with which we initialized the systems ($\alpha = 1.30$; \citealt{hwang_superthermal_22}) to nearly thermal within the first ${\sim}1.5$\,Gyr, somewhat shorter than the median ionization time of 2.6\,Gyr for the 40\% of binaries that are ionized at some point during the full 9\,Gyr simulation. After that, $\alpha$ shows little significant evolution ($\alpha = 1.06 \pm 0.05$ at 9\,Gyr\footnote{Error bars show the $1\sigma$ statistical uncertainty on the maximum-likelihood estimate of $\alpha$, $\sigma_\alpha = (\alpha + 1)/\sqrt{N}$, where $N$ is the number of binaries still bound at that time.
}). These numerical results extend the work of \citet{modak_hamilton_tide_23} and \citet{hamilton_modak_encounters_24}---who treat the Galactic tide and stellar encounters separately and find that each process drives initially superthermal distributions toward thermal---to the case in which both processes operate concurrently in the Solar neighborhood.

\section{Evolutionary Pathways for Individual Planetary Systems}  \label{sec:binary_plus_planet}

In this Section we describe the qualitative individual outcomes that arise in our simulations that include planets, and we provide some illustrative examples. These simulations come from our grid that includes both the Galactic tide and stellar encounters. The masses and initial semimajor axes of each example system are listed within the corresponding figure.

\subsection{Propagation of Eccentricity Excitation} \label{sec:planets_passages}

For binaries with $a_\mathrm{b} \gtrsim 10^4$\,AU, the companion star is too distant to perturb the planet appreciably except during brief phases of the binary's evolution. For our typical system parameters, the quadrupole ZLK timescale \citep{antognini_2015} is $\gtrsim 20$\,Gyr at $a_\mathrm{b} = 10^4$\,AU, which exceeds the age of the Universe. The planet is strongly perturbed only when the Galactic tide and stellar encounters drive the binary's orbit to high eccentricity, which recurs on the tidal oscillation timescale (Equation~\ref{eq:tsec}). Because the ZLK timescale falls steeply with the binary's pericenter distance ($t_{\rm ZLK} \propto [a_\mathrm{b}^2(1-e_\mathrm{b})]^{3/2}$), a binary that lingers at pericenter distances of tens to hundreds of AU for longer than $t_{\rm ZLK}$ can secularly excite the planet's eccentricity, while a single pericenter dip that brings the secondary star within a few tens of AU of the planet can change the planet's orbit impulsively. This process is similar to mechanisms studied by \cite{zakamska2004excitation}, \cite{kaib_2013}, \cite{yang2025third}, and \cite{yang2026analytic} among others.

\begin{figure*}
    \centering
    \includegraphics[width=1.0\linewidth]{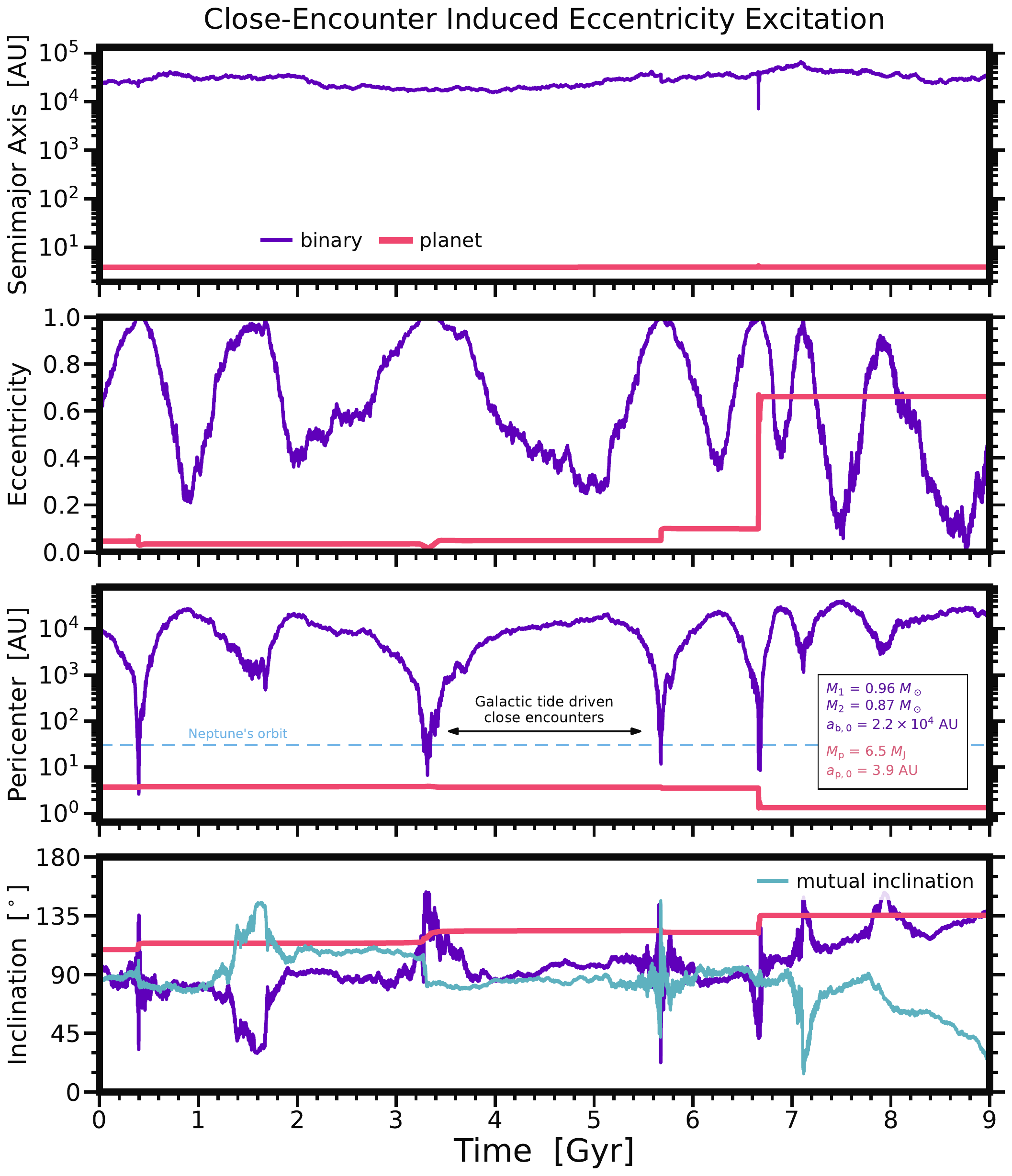}
    \caption{Evolution of a system in which the binary is driven to near-unity eccentricities, leading to close encounters with the planet that dramatically excite its eccentricity. Panels show, from top to bottom, the semimajor axis, eccentricity, pericenter distance, and inclination of both the binary (purple) and the planet (pink). The planet--binary mutual inclination is plotted in teal in the bottom panel. The orbital elements are calculated in Jacobi coordinates. Inclinations of the individual orbits are measured relative to the Galactic plane, and the system's parameters are listed in an annotation in the third panel. For scale, the dashed blue line in the third panel marks Neptune's orbital separation.}
    \label{fig:eccentricity_excitation_example}
\end{figure*}

An example of this evolution is shown in Figure~\ref{fig:eccentricity_excitation_example}. Over the 9\,Gyr integration, the binary's orbital evolution includes four deep dips in the binary pericenter to $a_\mathrm{b}(1-e_\mathrm{b})<100$ AU, at $t \approx 0.4$, 3.3, 5.7, and 6.6\,Gyr. While the first three pericenter dips result in only minor perturbations to the planet's orbit, the fourth raises the planet's eccentricity from ${\sim} 0.1$ to ${\sim} 0.67$, where it remains for the rest of the integration. Such excitation, in which the planet's eccentricity departs from its initial value by more than 0.1 without subsequent ejection or disruption, occurs in ${\sim}9\%$ of our systems. In about two-thirds of these, the planet's eccentricity changes abruptly in a single impulsive event, while the remainder undergo a brief episode of 1-2 secular oscillations before settling at a higher eccentricity value.

\subsection{Planetary Ejections and Tidal Disruptions} \label{sec:planets_ejections}

Close encounters can also completely eject the planet or result in tidal disruptions. Over 9\,Gyr, ${\sim}10\%$ of our planets are either ejected or disrupted. We examine how these losses depend on the binary's properties and the system's age in Section~\ref{sec:pop_loss}, and we discuss the properties of the ejected planets in Section~\ref{subsec:ffp}.

\begin{figure}
    \centering
    \includegraphics[width=1.0\linewidth]{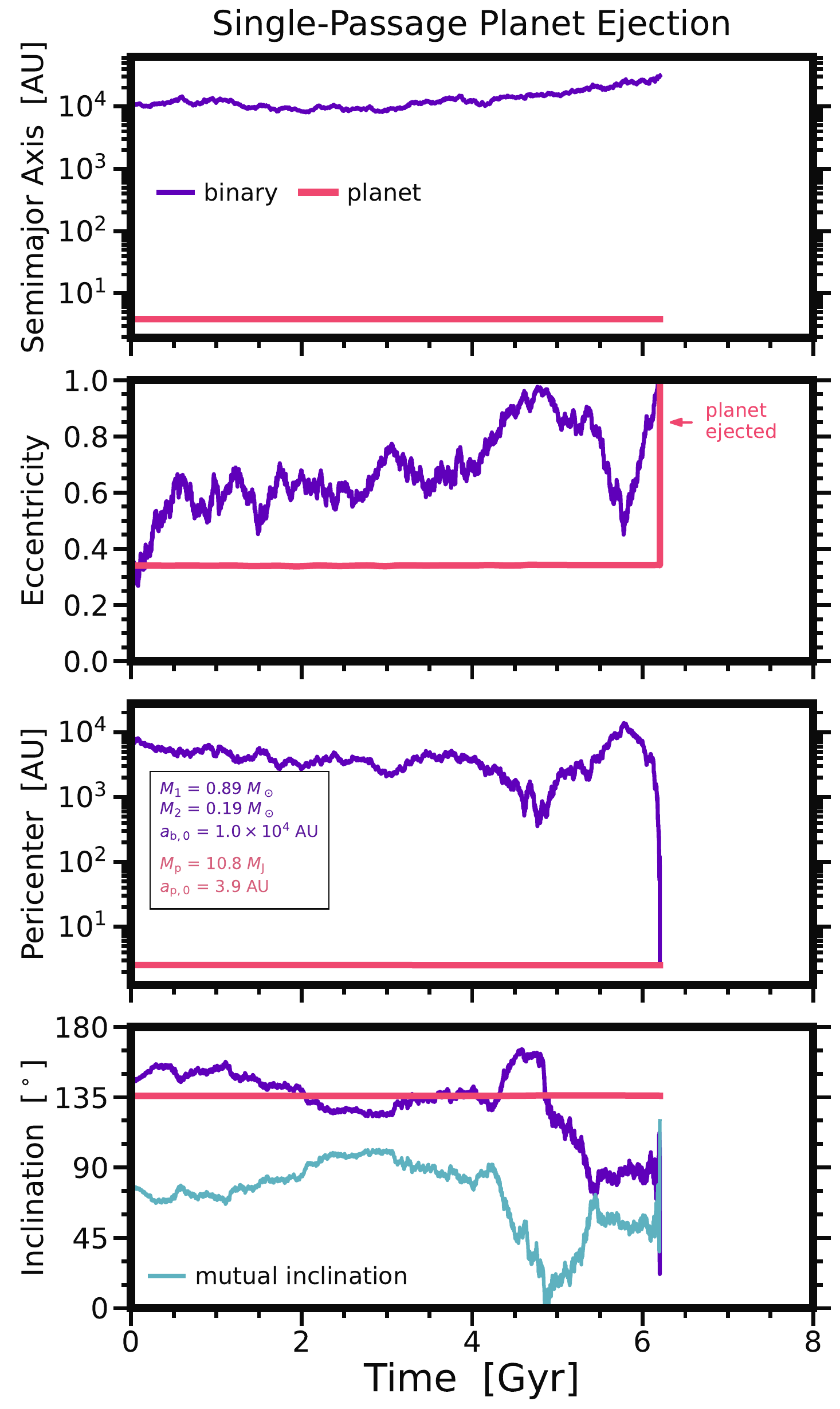}
    \caption{Analogous to Figure~\ref{fig:eccentricity_excitation_example}, but for a planet ejected in a single close passage. This is the more common path to ejection, by comparison with secular excitation followed by ejection (Figure \ref{fig:slow_ejection_example}). The planet's orbit is essentially unchanged for over 6\,Gyr until the binary's pericenter plunges from ${\sim}10^4$\,AU to ${\sim}3$\,AU within $\sim$20\,Myr, and the planet is ejected at $t \approx 6.2$\,Gyr with minimal prior orbital evolution.}
    \label{fig:fast_ejection_example}
\end{figure}

Ejections are typically abrupt: half of the ejected planets are lost within a few Myr of their first $\Delta e_\mathrm{p} > 0.1$ eccentricity excitation, and ${\sim}70\%$ are lost within 100\,Myr. This indicates that a single close passage by the binary companion usually suffices to trigger the planet's ejection, as illustrated by Figure~\ref{fig:fast_ejection_example}. 

\begin{figure}
    \centering
    \includegraphics[width=1.0\linewidth]{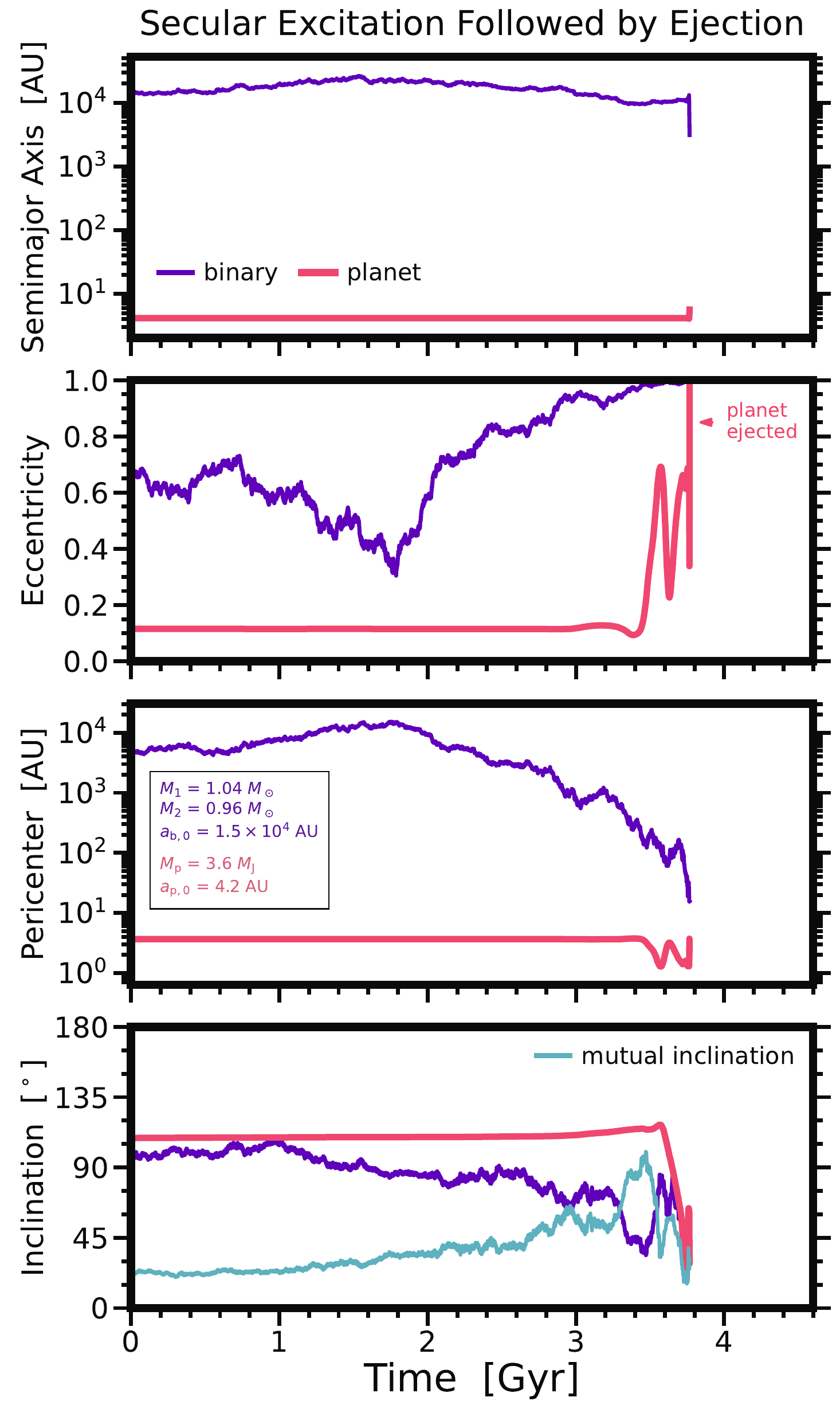}
    \caption{Analogous to Figure~\ref{fig:eccentricity_excitation_example}, but for a system in which the planet is ejected after a prolonged phase of secular excitation. The binary's eccentricity grows steadily over $\sim$2\,Gyr, bringing its pericenter down from ${\sim}10^4$ AU to a few hundred AU, close enough for the companion to drive secular oscillations in the planet's eccentricity (up to ${\sim}0.7$) while its semimajor axis stays fixed. Over the final ${\sim}$7\,Myr, a sequence of close passages at orbital distances ${\sim} 15$--$30$\,AU exchanges energy between the two orbits, hardening the binary and loosening the planet, until the planet is ejected at $t\approx 3.8$\,Gyr.}

    \label{fig:slow_ejection_example}
\end{figure}

In some cases, the planet may undergo secular oscillations induced by a close approach with the binary companion before it is ejected from the system. Figure~\ref{fig:slow_ejection_example} shows a case in which the binary lingers at pericenter distances of a few hundred AU, driving growing secular oscillations in the planet's eccentricity for $\sim$0.3\,Gyr before a final sequence of close passages over $\sim$7\,Myr ejects the planet from the system.

Close passages and secular excitation can also on occasion drive the planet's periapsis within the Roche limit of its host star (Equation~\ref{eq:roche}), resulting in tidal disruption of the planet. Since our simulations do not account for stellar or planetary tides, we cannot distinguish between cases in which the planet is engulfed or instead becomes a hot Jupiter. It is conceivable that a substantial fraction of such cases would culminate in hot Jupiter formation, as discussed by \cite{grishin_wide_binary_hj_25}. We estimate hot Jupiter formation rates for our simulations in Section~\ref{subsec:tides}.

\subsection{Binary Ionization by the Planet}\label{sec:planets_ionize}

In some cases, a close encounter between the stellar companion and the planet can ionize the binary while leaving the planet still bound to its star. These events can be understood as a consequence of the relative binding energies of the planet's and binary's orbits. 

\begin{figure}
    \centering
    \includegraphics[width=1.0\linewidth]{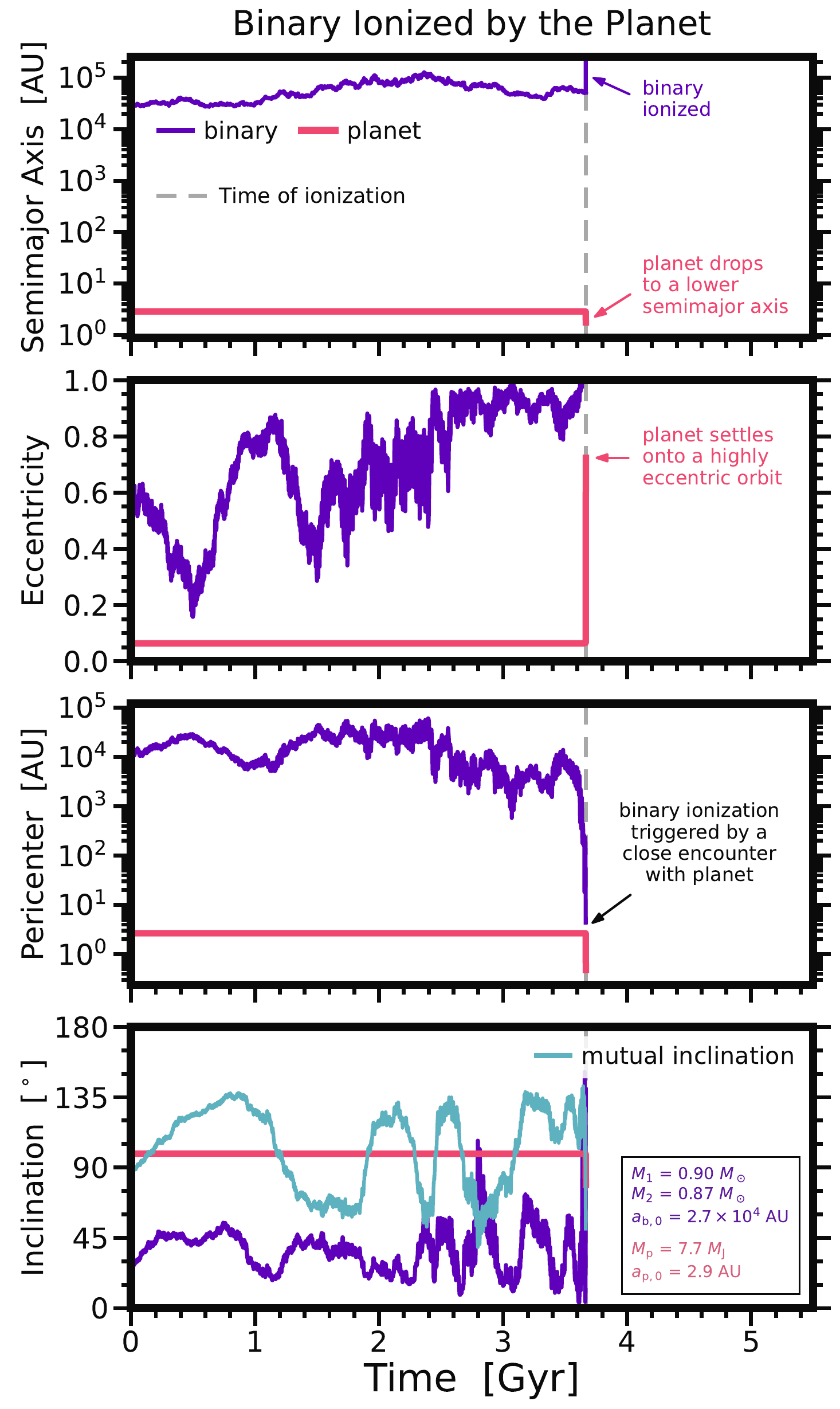}
    \caption{Analogous to Figure~\ref{fig:eccentricity_excitation_example}, but for a system in which the planet ionizes the binary. After several Gyr of high eccentricity excited by stellar encounters, the binary's pericenter drops to $r_p \approx 4$\,AU, comparable to the planet's orbit. In the resulting close encounter, the planet gives up orbital energy to the companion, unbinding the binary at $t \approx 3.7$\,Gyr (vertical dashed line) while its own semimajor axis shrinks from $2.9$ to $1.7$\,AU and its eccentricity jumps to $e_\mathrm{p} \approx 0.7$.}

    \label{fig:binary_ionization_example}
\end{figure}

To illustrate this point: an $8\,M_\mathrm{J}$ super-Jupiter on a 3\,AU orbit around a solar-mass star has a binding energy of $E_\mathrm{p} = G M_\odot M_\mathrm{p} / 2a_\mathrm{p} \approx 2 \times 10^{43}$\,ergs, while a binary comprised of two solar-mass stars with orbital separation $10^4$\,AU has a binding energy of only $E_\mathrm{b} = G M_\odot^2 / 2a_\mathrm{b} \approx 9 \times 10^{41}$\,ergs, roughly 25 times smaller. A close encounter between the planet and stellar companion can therefore transfer enough orbital energy from the planet to the binary to unbind the secondary star while leaving the planet's orbit bound, though substantially altered. Figure~\ref{fig:binary_ionization_example} shows an example of this scenario. A deep pericenter dip in the binary orbit at $t \approx 3.7$\,Gyr leads to a close encounter that ionizes the binary, moving the planet to a smaller semimajor axis and leaving it on a highly eccentric orbit. Across our grids, planet-hosting simulations produce ${\sim}5\%$ more ionized systems than planet-free simulations (see Section~\ref{sec:planet_ionization} for a more detailed discussion).

\subsection{Secular Evolution in Tighter Binaries: Modulated ZLK}\label{sec:planets_secular}

For binaries toward the lower end of our examined initial separations ($a_\mathrm{b} \approx 10^3$\,AU), the companion is close enough to drive ZLK-like oscillations in the planet's eccentricity and inclination over the examined timescales. For a fixed binary orbit, these oscillations would be regular and periodic. However, the binary's orbit is not fixed: at these separations it is perturbed significantly by stellar encounters, which dominate over the Galactic tide in their overall influence for binaries with $a_\mathrm{b} \lesssim 10^4\,$AU \citep{hamilton_modak_encounters_24}. Consequently, the period and amplitude of the planet's oscillations change along with the binary's orbit, and the planet undergoes a modulated version of ZLK evolution. This modulated ZLK evolution occurs in ${\sim}18\%$ of systems, about two-thirds of which are binaries with initial separations $a_{\rm b,0} \lesssim3\times10^3$\,AU.

\begin{figure}
    \centering
    \includegraphics[width=1.0\linewidth]{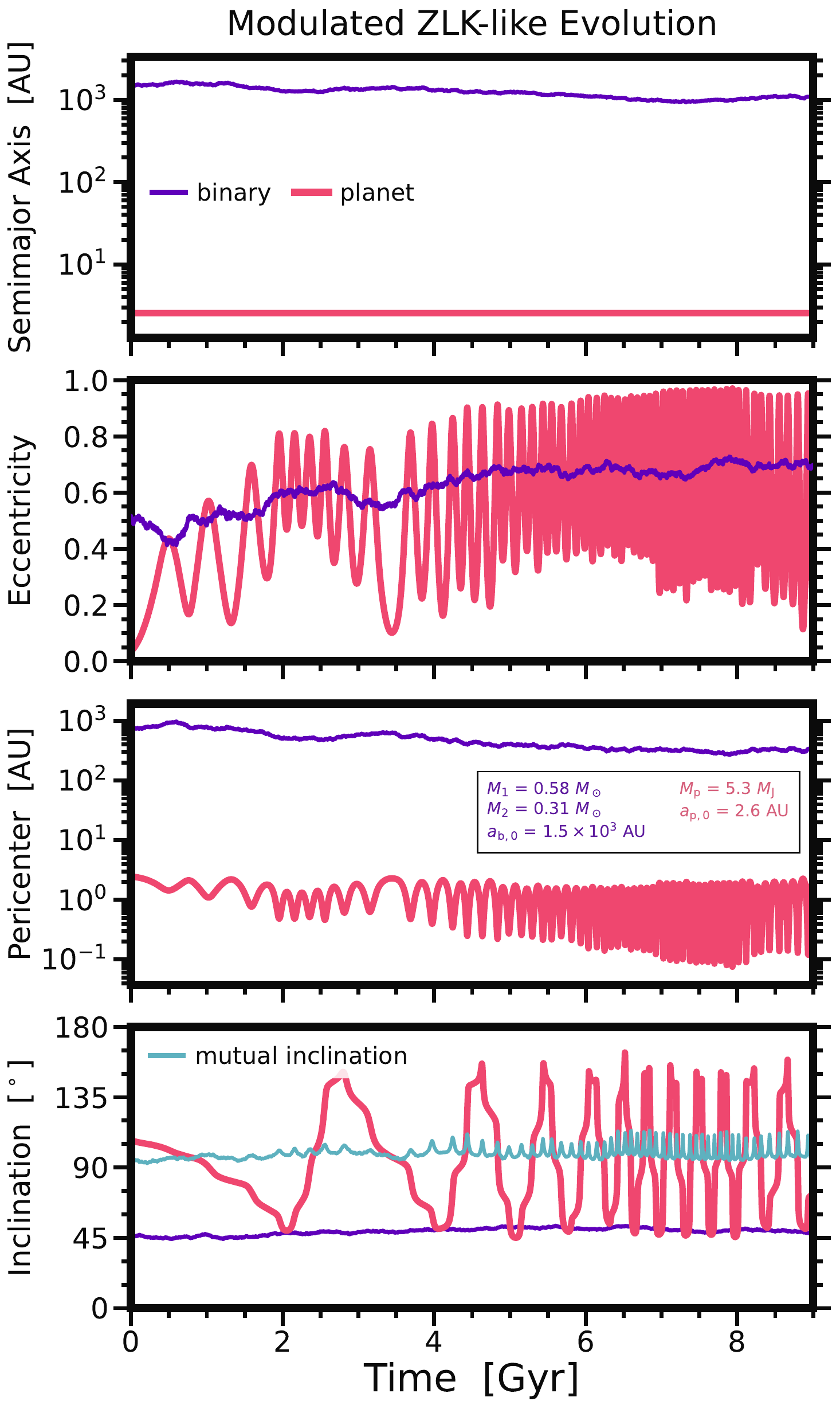}
    \caption{Analogous to Figure~\ref{fig:eccentricity_excitation_example}, but for a relatively tight binary ($a_{\rm b} \approx 10^3$\,AU) whose orbit is initially nearly perpendicular to the planet's ($i_{\rm mut} \approx 94.9^\circ$). The companion drives ZLK-like oscillations in the planet's eccentricity, reaching $e_\mathrm{p} \approx 0.97$ (pericenter $\approx 0.08$\,AU). These oscillations are modulated by stellar encounters, which continually perturb the binary's orbit, gradually shrinking it and raising its eccentricity. As a result, the oscillations speed up roughly fourfold over 9\,Gyr, from an oscillation period of ${\sim}400$ Myr to ${\sim}100$\,Myr.}
    \label{fig:kozai_like_example}
\end{figure}

Figure~\ref{fig:kozai_like_example} shows an example of this behavior. The planet oscillates between a nearly circular orbit and $e_\mathrm{p} \approx 0.97$ throughout the integration, with the oscillations becoming more rapid as stellar encounters gradually shrink the binary and raise its eccentricity. Over 9\,Gyr, the period of the ZLK oscillations experienced by the planet shortens roughly fourfold. At its most eccentric orbit, the planet's pericenter reaches roughly $0.08$\,AU. This is close enough to the host star that tidal dissipation in the planet, which we do not model, would become important in following the subsequent evolution (see Section~\ref{subsec:tides} for further discussion).

The binary's orbit changes appreciably only over Gyr timescales. For processes that complete within a few cycles, such as rapid high-eccentricity migration over tens to hundreds of Myr \citep[e.g.,][]{vick2019chaotic}, a fixed binary orbit is therefore a reasonable approximation. Over Gyr timescales, however, the cumulative effect of encounters is substantial: among binaries with $a_\mathrm{b} < 3\times10^3$\,AU, including encounters nearly doubles the fraction of planets driven to periapses below 0.05\,AU over 9\,Gyr (14\% versus 8\% with the tide alone) and nearly triples the fraction tidally disrupted (11\% versus 4\%). Stellar encounters should therefore be accounted for when following the long-term ZLK-like evolution of planets in binaries with separations of a few thousand AU.

\subsection{Quiescent Evolution}\label{sec:planets_quiet}

\begin{figure}
    \centering
    \includegraphics[width=1.0\linewidth]{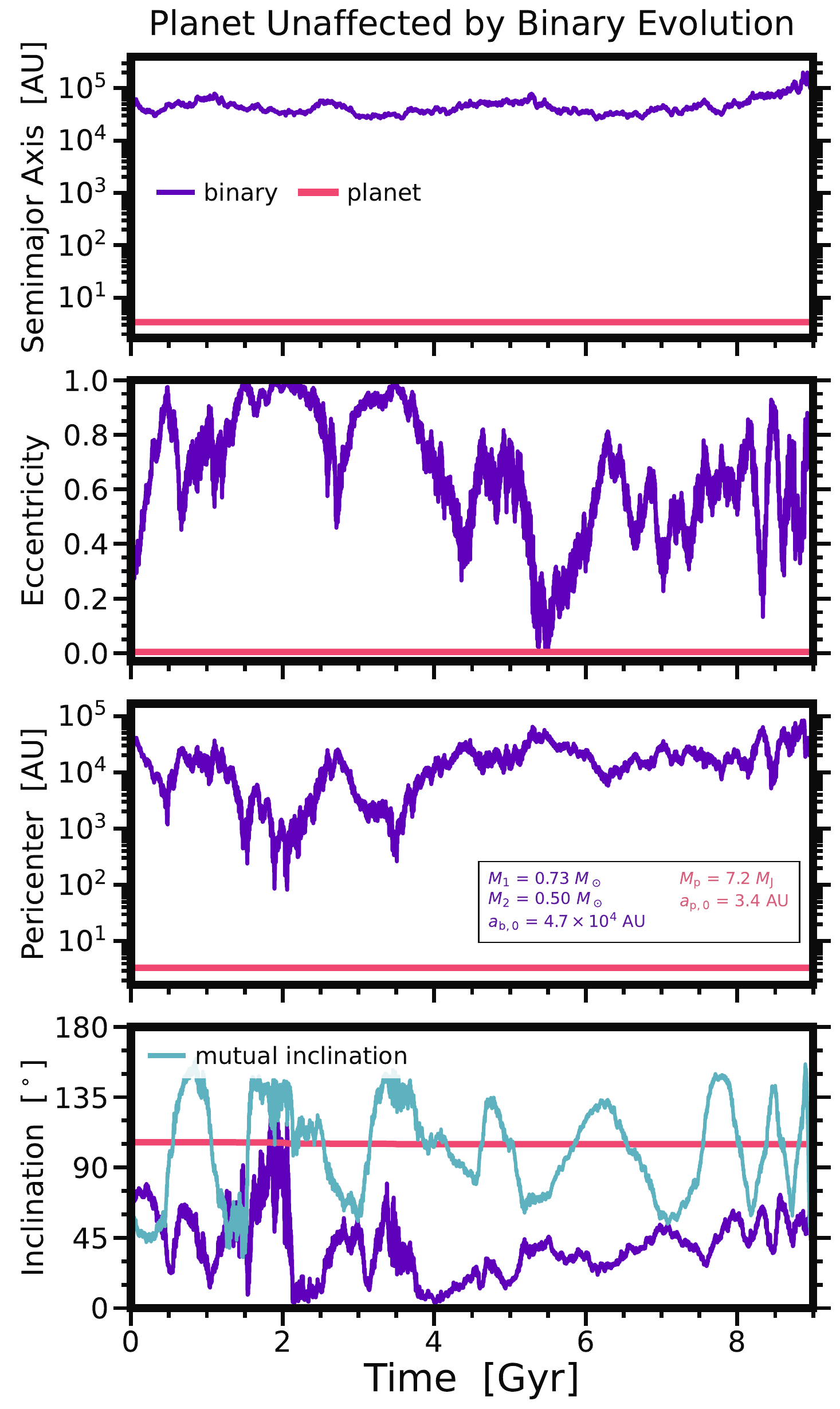}
    \caption{Analogous to Figure~\ref{fig:eccentricity_excitation_example}, but for a wide binary ($a_{\rm b} \approx 5 \times 10^4$\,AU) whose evolution leaves the planet unaffected. The binary's eccentricity and inclination evolve significantly under the Galactic tide and stellar encounters, but its pericenter never drops below ${\sim}80$\,AU, and the ZLK timescale at this separation far exceeds 9\,Gyr. The planet's orbit is therefore essentially unchanged; the variations in mutual inclination reflect the reorientation of the binary, not of the planet.}

    \label{fig:quiet_planet_example}
\end{figure}

In many cases, the planet's orbit remains relatively unperturbed over the 9 Gyr integration. Often, the cumulative effect of stellar encounters ionizes the binary without ever driving it to sufficiently small pericenter approaches to significantly influence the planet's orbit. In other cases, particularly at low initial Galactic inclinations, the eccentricity oscillations of the binary remain low-amplitude over the full integration. In both cases, the binary companion does not significantly perturb the planet. Quiescent evolution is the most common outcome in our grid of simulations, accounting for ${\sim}60\%$ of systems. In 57\% of these (905 of 1579), the binary is ionized without having significantly perturbed the planet.

Figure~\ref{fig:quiet_planet_example} shows one example of binary evolution that does not significantly affect the planet. Here, the binary experiences large oscillations in eccentricity and Galactic inclination over 9\,Gyr, but its pericenter never drops below ${\sim} 80$\,AU, roughly 25 times the planet's semimajor axis. The planet's orbit remains essentially unchanged over the 9 Gyr integration. 

Comparing with Figures~\ref{fig:eccentricity_excitation_example},  \ref{fig:fast_ejection_example} and \ref{fig:slow_ejection_example}, we see that the planet's response depends strongly on the geometry of the companion's approach: most importantly its pericenter distance, but also the relative inclination of the two orbits. In the next section, we quantify how often each of these outcomes occurs across our full population of simulations and how they depend on the properties of the binary.

\section{Population-Level Results} \label{sec:planet_population}

In this section, we examine our results to constrain the architectures of super-Jupiters in wide binaries on a population level. While this Section encompasses all three simulation grids, unless otherwise specified results are reported from our grid including both the Galactic tide and stellar encounters. We offer predictions for upcoming observations from Gaia DR4 and DR5.

\subsection{Evolution of Eccentricities}\label{sec:pop_ecc}

\begin{figure*}
    \centering
    \includegraphics[width=1.0\linewidth]{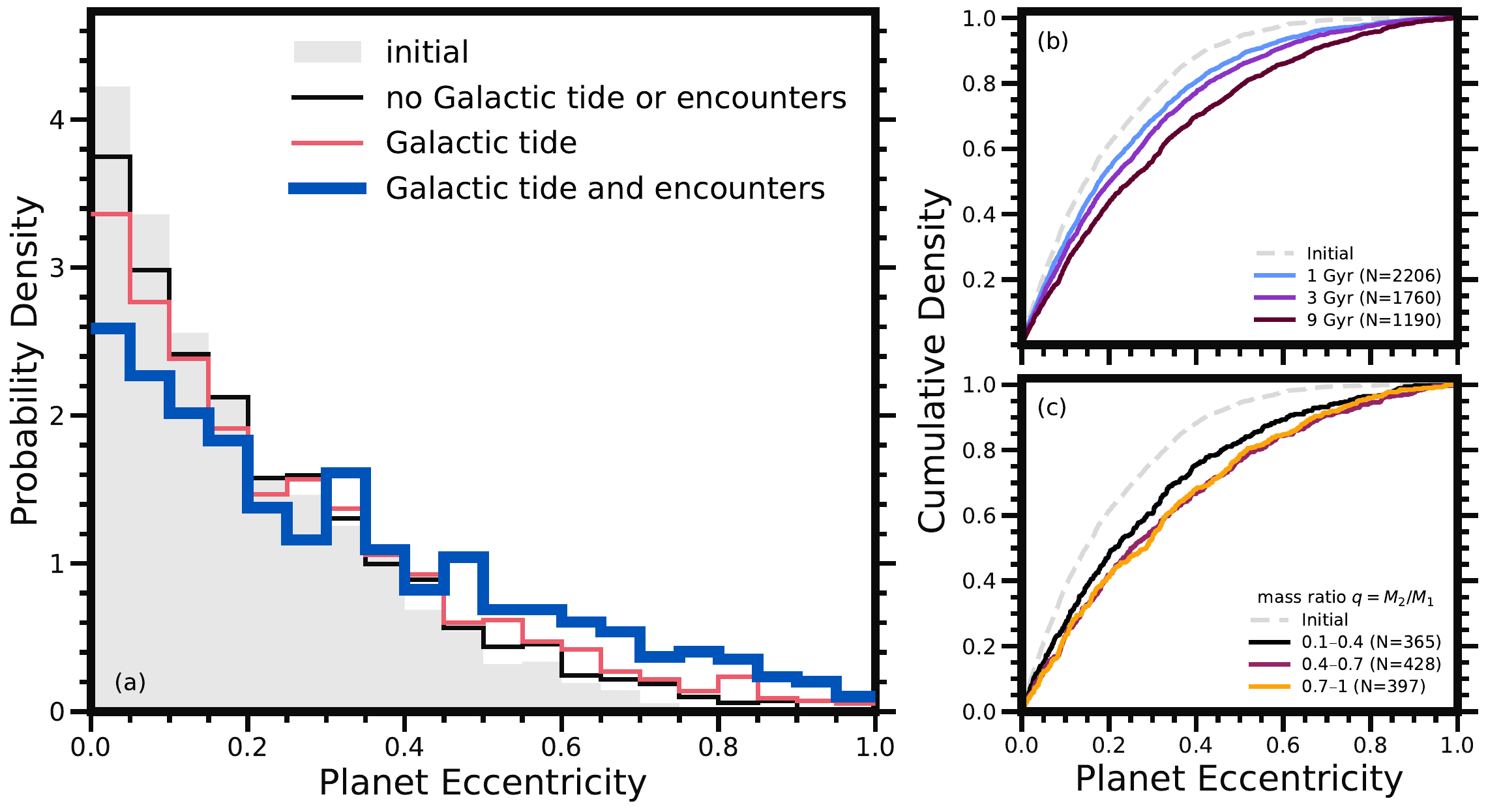}
    \caption{Eccentricity distributions of planets in surviving wide binaries in which both the planet and the binary companion remain bound. \textit{(a):} Distributions after 9\,Gyr for our grids with neither the Galactic tide nor stellar encounters (black), the Galactic tide alone (red), and both (blue), compared to the initial distribution (grey). \textit{(b):} Cumulative distributions for the grid that incorporates the Galactic tide and stellar encounters, displayed at 1, 3, and 9\,Gyr. \textit{(c):} Cumulative distributions at 9\,Gyr for the grid that incorporates the Galactic tide and stellar encounters, split by binary mass ratio $q = M_2/M_1$. The number of surviving systems in each subsample is given in the legends. The Galactic tide and stellar encounters together substantially excite planetary eccentricities, with the excitation growing over time and the strongest eccentricity excitation observed for planets orbiting stars with more massive companions.}
    \label{fig:ecc_evolution}
\end{figure*}

Figure~\ref{fig:ecc_evolution} summarizes the evolution of the eccentricity distribution of planets in surviving wide binaries. Panel (a) compares the final distributions across our grids. Even in the absence of both the Galactic tide and stellar encounters, the distribution evolves to systematically higher eccentricities from its initial state due to ZLK oscillations (Section~\ref{sec:planets_secular}). Both the Galactic tide and stellar encounters act to further push the distribution toward even higher eccentricities.

This eccentricity excitation builds up gradually over time (panel b). With both the tide and encounters, the median planetary orbital eccentricity rises from ${\sim}0.15$ initially to ${\sim}0.18$ after 1 Gyr, ${\sim}0.20$ after 3 Gyr, and ${\sim}0.25$ after 9 Gyr. We thus predict that giant planets in wide binaries should be more eccentric in older systems. The number of surviving planets in surviving binaries also declines over this timespan, from 2206 at 1\,Gyr to 1190 at 9\,Gyr. This decline primarily stems from the ionization of the binaries themselves (Sections \ref{sec:binary_pop} and \ref{sec:planets_ionize}), but it is also influenced by planet ejections and disruptions (Section \ref{sec:planets_ejections}). The planets in ionized binaries, some of which were excited to higher-eccentricity orbits before their companion was lost, would orbit single stars today (see Section \ref{sec:single_stars} for further discussion of this outcome).

The degree of eccentricity excitation also depends on the binary's mass ratio (panel c). Planets hosted by stars with relatively low-mass stellar companions ($q = 0.1$--$0.4$) are noticeably less excited than those with more massive stellar companions, consistent with the weaker perturbation that a lower-mass companion exerts during each close passage. Above $q \approx 0.4$, however, the dependence on mass ratio largely disappears: the distributions for $q = 0.4$--$0.7$ and $q = 0.7$--$1$ are statistically indistinguishable
(Kolmogorov--Smirnov $p = 0.93$).

Taken together, our results predict that the eccentricity distribution of super-Jupiters detected in wide binaries by Gaia DR4 and DR5 should be excited relative to that of planets around single stars, with the eccentricity enhancement strongest in older systems and in binaries with more massive companions.

To estimate whether Gaia will be able to detect this eccentricity excitation, we construct a mock sample of planets in wide binaries from our simulations, observed at ages drawn uniformly from 0--9\,Gyr (to simulate a constant star formation rate), including only systems in which both the planet and the companion remain bound. To mimic realistic Gaia measurement errors, we draw them empirically from the \citet{lammers_winn_gaia_26} DR5 mock catalog: each simulated planet is assigned the best-fit eccentricity of a randomly chosen planet among the 50 catalog planets closest to it in true eccentricity (typically within ${\sim}10^{-2}$), thereby inheriting the measurement uncertainties of a planet with essentially the same eccentricity. We then compare the resulting sample to planets drawn from the catalog itself, which represent the single-star planet population, using an Anderson--Darling test \citep{anderson_darling_1952, scholz_stephens_1987}. We find that distinguishing the two distributions at $p < 0.05$ in 90\% of trials requires $\gtrsim250$ planets in wide binaries, or $\gtrsim400$ for a $3\sigma$ detection.

Since the hosts in the \citet{lammers_winn_gaia_26} mock catalogs are real Gaia sources, we can count their actual wide companions directly: cross-matching with the wide-binary catalog of \citet{elbadry_catalog_21}, we find that 447 of the 7155 DR4 hosts (6.2\%) and 6283 of the 116{,}201 DR5 hosts (5.4\%) with planets beyond 1\,AU have a detected companion at projected separations of $10^3$--$10^5$\,AU with chance-alignment probabilities $<0.1$. DR4 may therefore already reveal this population-level eccentricity excitation, and DR5 should do so robustly.

The trends with system age and binary mass ratio rely on smaller differences between subsets of the wide-binary population, so a two-sample test on best-fit eccentricities would need larger samples to detect them: ${\sim}7000$ planets in wide binaries to distinguish systems younger and older than 4.5\,Gyr at $3\sigma$ (assuming perfectly known stellar ages), and ${\sim}16{,}000$ to separate binaries with $q < 0.4$ from those with $q \geq 0.4$. Radial-velocity follow-up, which can tighten individual eccentricity measurements, and more sophisticated population models, such as hierarchical Bayesian frameworks that infer population-level eccentricity distributions from the orbit posteriors of individual systems \citep[e.g.][]{hogg_2010,van_eylen_2019,bowler_blunt_nielsen_2020, do_o_23, nagpal_23, sagear_ballard_23, gilbert_smooth_transition_26}, may allow these trends to be detected with considerably smaller samples.

\subsection{Distributions in Inclination}\label{sec:pop_incl}

\begin{figure}
    \centering
    \includegraphics[width=1.0\linewidth]{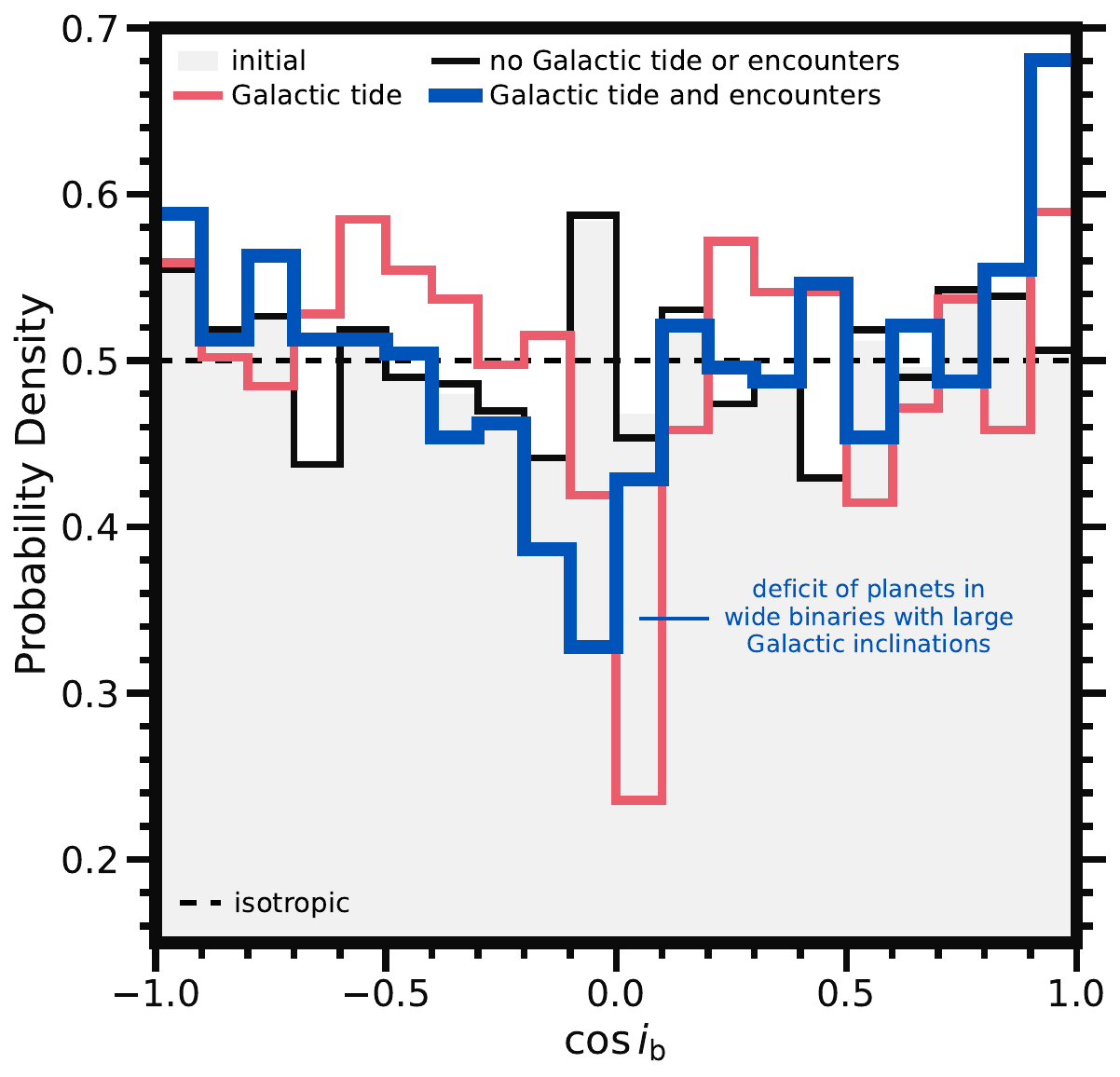}

    \caption{Distribution of binary Galactic inclinations for surviving planets in surviving wide binaries at 9\,Gyr. With both the Galactic tide and stellar encounters, surviving planets show a deficit near $\cos i_\mathrm{b} = 0$, i.e. in binaries whose orbits are nearly perpendicular to the Galactic midplane. The deficit grows with age and is weaker in younger systems.}
    \label{fig:planet_binary_inc_distribution}
\end{figure}

Since the Galactic tide acts most strongly on binaries with large Galactic inclinations (Section~\ref{sec:binary_only}), the dynamical processes discussed so far leave an imprint in the orientation distribution of planet-hosting wide binaries. Without the Galactic tide or stellar encounters, the distribution of Galactic inclinations for binaries that retain their planet after 9\,Gyr does not significantly evolve from isotropy (Figure~\ref{fig:planet_binary_inc_distribution}). When the Galactic tide is taken into consideration, however, a narrow but pronounced deficit arises, corresponding to an underdensity of binaries with orbital planes nearly perpendicular to the Galactic midplane. These are the low-$j_z$ systems that the Galactic tide drives to close pericenter passages, which often result in close approaches that destabilize planets. When stellar encounters are also included, this deficit is broadened and weakened as encounters allow binaries to diffuse into and out of the low-$j_z$ region (Section~\ref{sec:binary_kicks}). Nevertheless, a preference against nearly perpendicular binaries still persists.

This deficit is a consequence of planet loss rather than of the binaries' own orbital evolution: in our simulations without planets, the Galactic inclinations of wide binaries remain close to isotropic, owing to the effects of stellar encounters. As a result, the deficit appears only within the distribution of planet-hosting binaries.

Testing this prediction observationally will require constraining the three-dimensional orientations of wide-binary orbits, which may pose a considerable challenge. Line-of-sight separations of $\lesssim 10^4$\,AU produce parallax differences below Gaia's precision for all but the nearest systems, and line-of-sight orbital velocities of ${\sim}0.1$\,km\,s$^{-1}$ are expected to be smaller than the uncertainties of Gaia's radial velocities for most hosts.

Precise radial-velocity follow-up may supply some of this missing line-of-sight information. Alternatively, statistical approaches such as that of \citet{hwang_superthermal_22} could constrain the distribution of Galactic inclinations at the population level. Because the deficit is confined to nearly perpendicular orbits and develops over Gyr timescales (systems with $|\cos i_\mathrm{b}| < 0.1$ are underrepresented by ${\sim}25\%$ at 9\,Gyr, but by only ${\sim}10\%$ when averaged over ages of 1--9\,Gyr), such a test will require a large sample of planet-hosting wide binaries, such as those that DR5 is expected to reveal.

\subsection{Ejections and Disruptions}\label{sec:pop_loss}

Planetary ejections and disruptions each occur in our simulations. In the grid with both the Galactic tide and stellar encounters, ${\sim}6\%$ of planets are ejected (159/2500) and ${\sim}3\%$ are disrupted (85/2500) over 9\,Gyr, compared to ${\sim}3\%$ and ${\sim}1.4\%$ respectively with the Galactic tide alone, and ${\sim}0.4\%$ and ${\sim}0.8\%$ with neither. With the Galactic tide and/or stellar encounters, we thus see that disruptions are roughly half as common as ejections. With neither perturbation, disruptions would dominate: the planets lost in this case are driven into their host stars by ZLK oscillations in the tightest binaries, rather than being ejected by close approaches when the binary pericenter dips due to long-term evolution.

\begin{figure}
    \centering
    \includegraphics[width=1.0\linewidth]{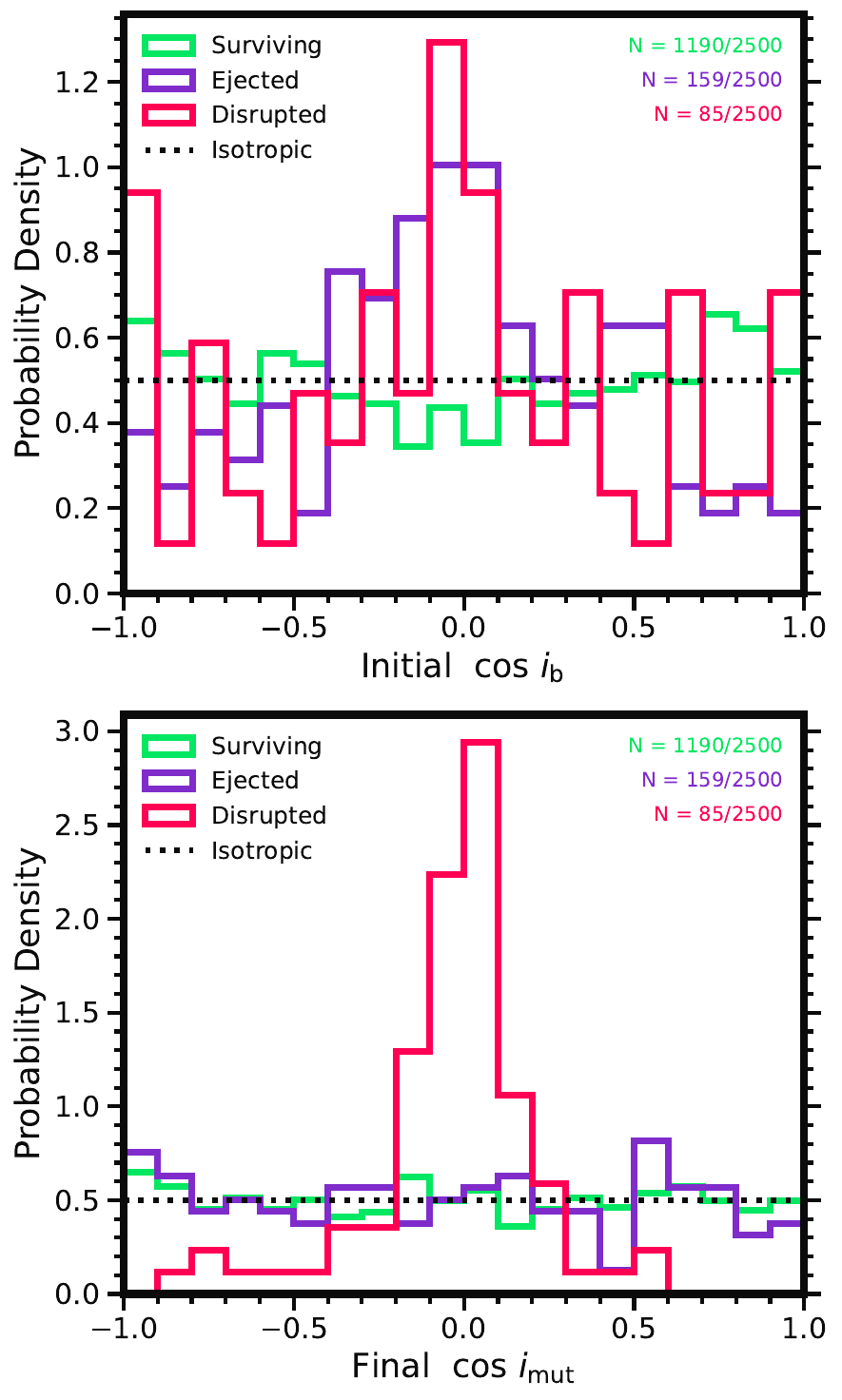}
    \caption{Inclination distributions by dynamical outcome of the planet. \textit{Top:} Initial binary inclination relative to the Galactic midplane. \textit{Bottom:} Final mutual inclination between the planet and binary, at 9\,Gyr for surviving planets and at the last bound snapshot for ejected and disrupted planets. Dotted lines denote isotropy. Ejected and disrupted planets preferentially derive from binaries with initial Galactic inclinations near $90^\circ$, where the tide most strongly excites the binary's eccentricity. Disrupted planets end near $90^\circ$ mutual inclination, consistent with expectations from octupole-order ZLK-driven dynamics (see Section \ref{sec:pop_loss}).}

\label{fig:galactic_inclination_vs_outcome}
\end{figure}

These outcomes depend strongly on the binary's initial Galactic inclination (Figure~\ref{fig:galactic_inclination_vs_outcome}, top panel). Ejected and disrupted planets preferentially occur in binaries with $\cos i_{\rm b,0} \approx 0$---that is, orbits initially nearly perpendicular to the Galactic midplane---while surviving planets are correspondingly depleted at $\cos i_{\rm b,0} \approx 0$. This reflects the dependence of the Galactic tide on Galactic inclination, as discussed in Section~\ref{sec:binary_tide}: only binaries with small $j_z$ can be driven to orbits with close pericenter passages needed to eject or disrupt the planet. This dependence is weaker than it would be under the Galactic tide alone, since stellar encounters allow binaries at other inclinations to reach low $j_z$ as well (Section~\ref{sec:binary_kicks}).

The final mutual inclinations between the planet and binary reveal another difference between the two outcomes (Figure~\ref{fig:galactic_inclination_vs_outcome}, bottom panel). For surviving planets, we report the final mutual inclination at 9\,Gyr. For ejected and disrupted planets, we instead report the mutual inclination at the last output ($\leq 0.8$\,Myr before the planet is lost) in which both the planet and the companion are still bound. The final mutual inclinations of surviving and ejected planets are both roughly isotropic, but those of disrupted planets are strongly concentrated near $90^\circ$. This suggests that the observed disruptions are driven primarily by octupole-order ZLK oscillations in the 3-body wide binary+planet system, which may produce simultaneously elevated mutual inclinations and eccentricities in the planet's orbit \citep[e.g.,][]{li2014eccentricity, naoz_araa}. Indeed, the systems with disrupted planets had systematically higher octupole coefficients \citep[Eq. 9 of][]{litchwick2011eccentric} than either the ejected or surviving populations. These disruptions are too few, however, to leave a significant imprint on the mutual inclinations of the surviving population, which remain roughly isotropic in all of our grids.

\begin{figure}
    \centering
    \includegraphics[width=1.0\linewidth]{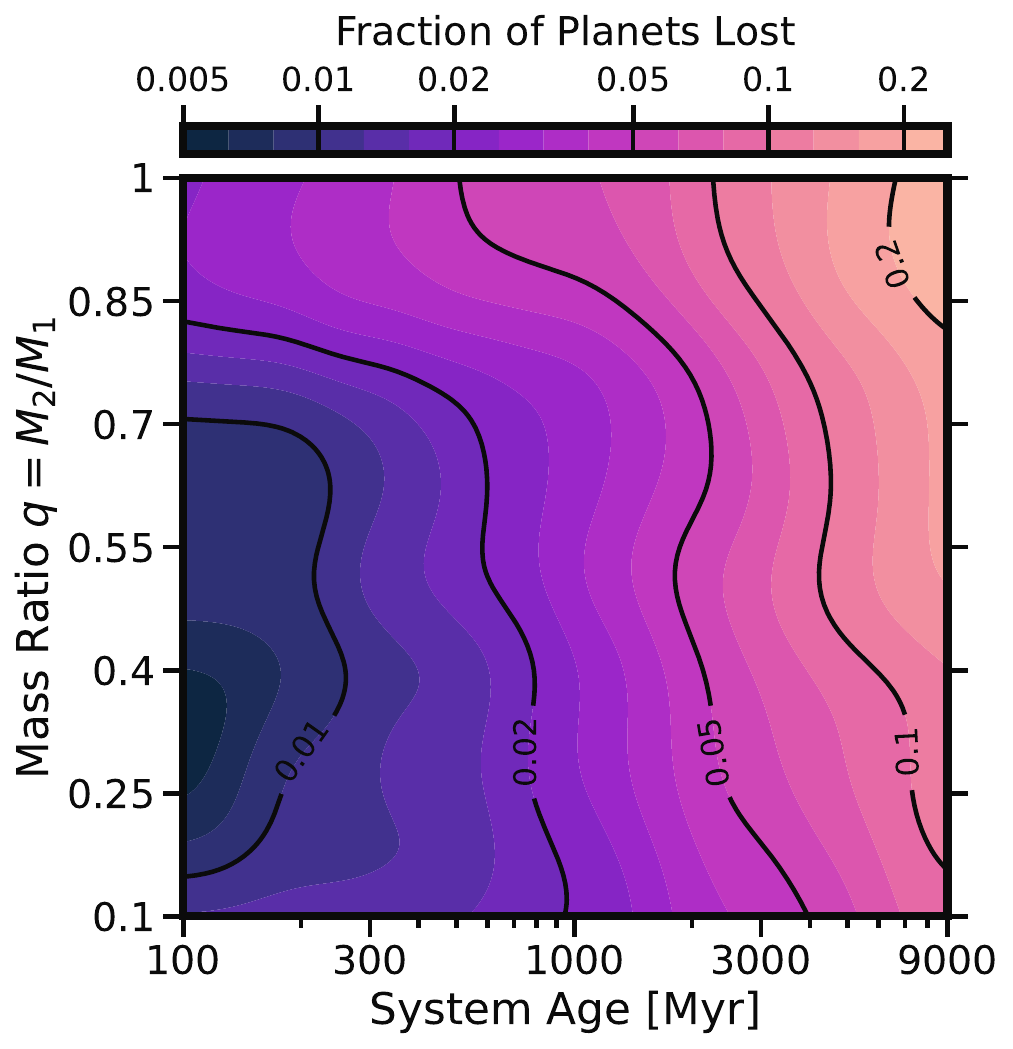}
    \caption{Cumulative fraction of planets lost (ejected or disrupted) as a
    function of system age and binary mass ratio $q = M_2/M_1$. Losses accumulate steadily with age and are most common in near-equal-mass binaries.}
    \label{fig:ejection_mass_ratio_age}
\end{figure}

Finally,  Figure~\ref{fig:ejection_mass_ratio_age} shows how the fraction of planets lost to either ejection or disruption depends on the binary's mass ratio and the system's age. These losses accumulate steadily over time, with most (70\%) occurring after the first Gyr. Ejections dominate this late-time growth: the fraction of planets ejected grows from 1.4\% at 1\,Gyr to 6.4\% at 9\,Gyr, while disruptions grow more modestly, from 1.5\% to 3.4\% over the same period. Losses are also substantially more common for near-equal-mass binaries: by 9\,Gyr, ${\sim}20\%$ of planets in binaries with $q > 0.775$ have been ejected or disrupted, compared to ${\sim}8\%$ for $q < 0.325$. This trend is driven mainly by ejections, which are nearly four times as common for $q > 0.775$ as for $q < 0.325$ (14\% versus 4\%), whereas disruptions vary only weakly with mass ratio (6\% versus 4\%). Two effects contribute: a more massive stellar companion perturbs the planet more strongly during each close pericenter passage, and more massive binaries are also more tightly bound (Equation~\ref{eq:tdis}), so they survive longer and undergo more close passages before being ionized. We hence predict that wide-orbiting giant planet occurrence should be increasingly depleted in older wide binaries and in those with more equal-mass components---a trend that may be testable with Gaia DR4 and DR5.

This trend should be especially pronounced for `twin' binaries. Gaia has revealed an excess of wide binaries with near-unity mass ratios \citep{kareem_twin_binaries_2019}, which also appear to be on significantly more eccentric orbits than the broader wide binary population that we model \citep{hwang_twin_binaries_22}. Since these systems combine near-equal stellar masses that efficiently lead to planet ejection with high eccentricities that naturally correspond to a higher occurrence of potentially destabilizing close encounters, our results suggest that wide-orbiting super-Jupiters may have a lower occurrence in Gyr-old twin wide binaries than in the broader population of binary star systems with lower-mass wide stellar companions. This, too, is a prediction that Gaia DR4 and DR5 could begin to test.

\subsection{Free-floating planets}
\label{subsec:ffp}

The planets ejected in our simulations will join the population of free-floating planets in the Galaxy. To investigate their properties, we re-initialize each simulation that results in a planet ejection, placing the system at its most recent checkpoint before the ejection event using \texttt{REBOUND}'s \texttt{SimulationArchive} \citep{rein2017simarchive} functionality. We then integrate each of these simulations forward through the ejection until the planet is well-separated from both stars ($10\times$ the binary's apocenter distance), with stellar encounters switched off after the ejection. We record the planet's asymptotic velocity relative to the binary's center of mass, $v_\infty$. Of the 159 planets that become unbound from their host star (Section~\ref{sec:pop_loss}), 5 are captured by the companion star and 2 remain bound to the binary as a whole, potentially becoming circumbinary planets\footnote{We note that we are unable to assess the long-term stability of any such-formed circumbinary planets, since the \texttt{TRACE} \texttt{WB} integrator is not designed to handle such cases.}. We exclude these 7 planets from the analysis in this section, leaving 152 free-floating planets that we examine in further detail.

\begin{figure*}
    \centering
    \includegraphics[width=1.0\linewidth]{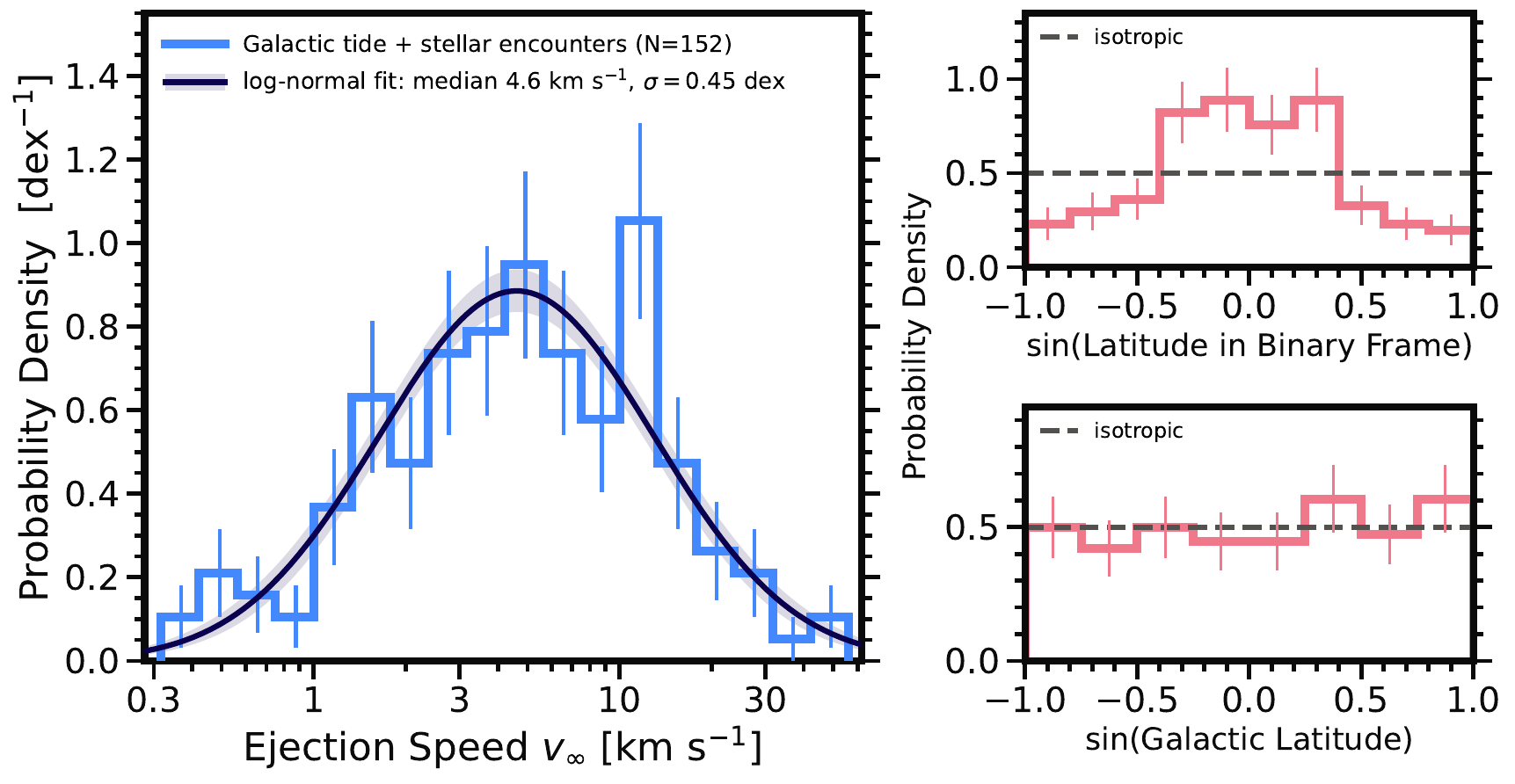} \caption{Properties of ejected planets, showing Poisson error bars on each panel. \textit{Left:} Distribution of asymptotic ejection speeds $v_\infty$ in blue, along with the best-fit log-normal distribution in black (median 4.6\,km\,s$^{-1}$, width 0.45\,dex). \textit{Top right:} Ejection directions relative to the binary's instantaneous orbital plane, expressed as the sine of their latitude in a frame whose equator is that plane. \textit{Bottom right:} Ejection directions in Galactic latitude. Dashed lines denote isotropy. Planets are preferentially ejected within the binary's orbital plane, which corresponds to a roughly isotropic distribution in Galactic coordinates. } \label{fig:ffp}
\end{figure*}

Figure~\ref{fig:ffp} summarizes the velocity and directional distributions of the ejected planets. Ejected planets span $v_\infty$ between $0.3$ and 50\,km\,s$^{-1}$, with a distribution that is well-described by a log-normal distribution with a median of 4.6\,km\,s$^{-1}$ and a width of 0.45\,dex. The ejection directions are concentrated towards the instantaneous orbital plane of the binary. Because the orbital planes of the binaries at the time of ejection are broadly distributed, the ejection directions appear effectively isotropic in Galactic coordinates.

Previous studies have demonstrated that planets may be ejected from Galactic perturbations to wide binaries, which may trigger planet-planet scattering in multi-planet systems \citep{kaib_2013} and lead to ejection or significant orbital evolution in single-giant-planet systems \citep{grishin_wide_binary_hj_25}. In the case that free-floating planets are ejected through planet-planet scattering, the mean excess speed has previously been shown to typically fall in the range of 2 km s$^{-1}$ $<v_{\infty}<6$ km s$^{-1}$ \citep{bhaskar2025properties}, in agreement with our results. Our ejection rates are lower than those of \citet{kaib_2013}, whose multi-planet systems offer additional avenues for instability through scattering. They are also lower than the ${\sim}20\%$ ejection rate found by \citet{grishin_wide_binary_hj_25}, likely because \citet{grishin_wide_binary_hj_25} considers planets at 5--200\,AU which are more easily ejected. Planets on wider orbits can be destabilized by far less extreme binary pericenter passages than the planets near the water-ice line that we consider. 

The free-floating planet population derived from wide binary ejections would likely be difficult to distinguish from the population that derives from single-star planet-planet scattering or binary-induced instabilities, as the typical projected ejection velocities and directions are comparable for each population \citep{rice2026dynamics}. Thus, in practice, the planets ejected through this framework may ``blend in'' with the distribution of planets ejected from more prevalent close binaries and single-star systems. 

\subsection{Planet-Driven Binary Ionization}
\label{sec:planet_ionization}

The mechanism of Section~\ref{sec:planets_ionize} requires the Galactic tide or stellar encounters to first drive the binary to $e_\mathrm{b} \to 1$: without either, only 1 of 2500 binaries in our simulations is ionized over 9\,Gyr ($<0.2\%$ at 95\% confidence). We first quantify how often the planet ionizes its binary under the Galactic tide alone by comparing each of our tide-only simulations with its counterpart in the binary-only grid (Section~\ref{sec:ics}), which shares identical initial conditions and the same Galactic tide but contains no planet. Excluding systems whose planet is ejected or disrupted first, we find that the planet is responsible for nearly all ionizations (108 with the planet, versus 13 without). These planet-driven ionizations occur throughout the 9\,Gyr (median 2.6\,Gyr), and none of the planet-free counterparts of the 99 systems that lose (either eject or disrupt) their planet are ionized. 

We next include stellar encounters, which are drawn from the same random sequence for each simulation and its binary-only counterpart. Individual outcomes are chaotic, but the net effect of the planet is clear: excluding systems whose planet is ejected or disrupted, 1066 binaries are ionized with the planet, compared with 941 of their planet-free counterparts. The planet therefore causes ${\sim}125$ additional ionizations, equivalent to $5\%$ of all 2500 systems, or a ${\sim}13\%$ increase over the number of binaries ionized without a planet. Stellar encounters also continue to unbind some binaries after their planet is lost: 26\% of the planet-free counterparts of systems whose planet is ejected or disrupted are eventually ionized. Because the pericenter passage that unbinds the companion also perturbs the planet, these planets are more often left on eccentric orbits around their now-single hosts: of the binaries ionized only in the presence of the planet, $24\%$ leave the planet with $e_\mathrm{p} > 0.5$, compared with $9\%$ of those whose planet-free counterparts are also ionized. We note that this channel is strengthened by the super-Jupiter masses of the planets in our simulations (median $8\,M_\mathrm{J}$). Due to the relatively low overall occurrence of such massive planets, this channel is unlikely to be a major contributor to the overall ionization rate of field wide binaries.

\subsection{Planets Around Former Wide Binary Components}\label{sec:single_stars}

In nearly half of our simulations, the binary is ionized while the planet remains bound to its host star. With both the Galactic tide and stellar encounters included, this outcome arises in 1066 of our 2500 systems (43\%), with a median ionization time of 2.5\,Gyr (10th--90th percentiles of 0.5--6.8\,Gyr). Planetary systems that have previously undergone this evolutionary pathway would orbit single stars, without necessarily showing clear imprints from the previously bound binary companion. Once the companion is lost, the planet's orbit is essentially frozen: stellar encounters kick the host star and planet identically (Section~\ref{sec:kicks}), and the Galactic tide's influence is negligible across an orbit of a few AU.

We report each planet's orbit at the final output of its simulation, which is the first output at which the binary meets our criteria for ionization. By this point, the companion star has receded to a median distance of $2.4\times10^5$\,AU from the planet's host, and it no longer significantly perturbs the planet. We report this output, rather than that of the last timestep in which the binary is still bound, because the ionizing passage itself can significantly alter the planet's orbit: in 11\% of these systems, the passage that ionizes the binary changes the planet's eccentricity by more than 0.05.

\begin{figure}
    \centering
    \includegraphics[width=1.0\linewidth]{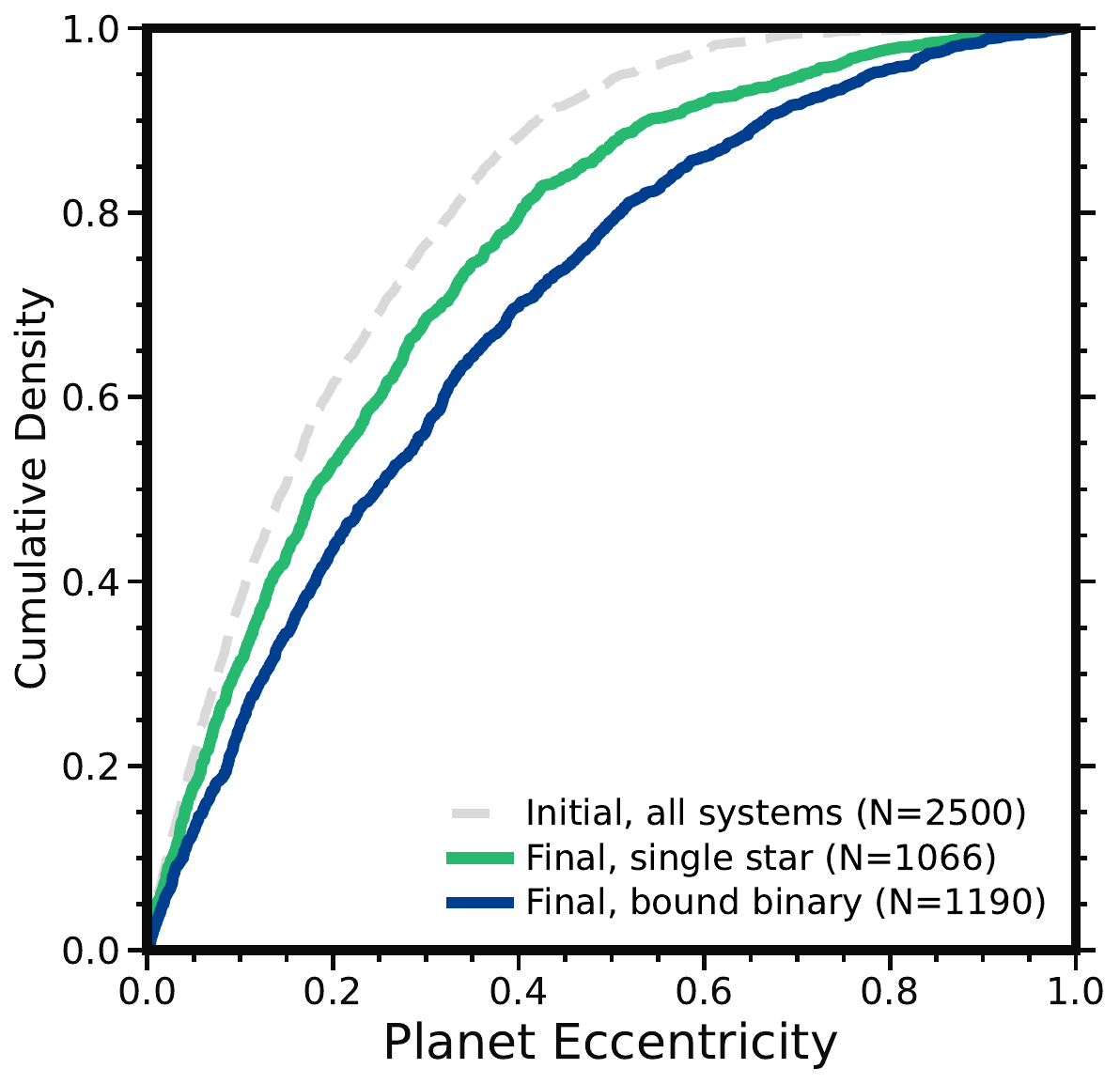}
    \caption{Cumulative eccentricity distributions of planets: the initial population (all 2500 systems), planets whose binary is ionized within 9\,Gyr, measured after the companion has departed ($N = 1066$), and planets in binaries that survive to 9\,Gyr ($N = 1190$). Planets left orbiting single stars are only mildly more eccentric than they began, and far less excited than those whose companion star remains bound, because the binaries that are ionized are predominantly the widest (see Equation~\ref{eq:tdis}).}
    \label{fig:single_stars}
\end{figure}

Figure~\ref{fig:single_stars} compares the eccentricities of these planets to those of the initial population and of planets in wide binaries that survive to 9\,Gyr. We find that planets around former wide binaries show only mild eccentricity excitation, with a median eccentricity of 0.18, compared to a median eccentricity of 0.15 in the initial distribution and 0.25 in surviving wide binaries. 13\% of planets in former wide binary systems have $e_\mathrm{p} > 0.5$, compared to 6\% initially and 21\% in surviving wide binaries. These differences arise because the ionized binaries are predominantly the widest ones (median initial separation of $3.6\times10^4$\,AU, compared to $3.8\times10^3$\,AU for survivors; Equation~\ref{eq:tdis}), which tend to be ionized relatively early, before many close encounters have occurred due to the Galactic tide and stellar flybys. This also implies an age dependence: 16\% of planets whose binary is ionized after 3\,Gyr have $e_\mathrm{p} > 0.5$, compared to 10\% for those ionized earlier.

Among the planets around former wide binary components, the 140 whose eccentricities increased by $\Delta e_\mathrm{p} > 0.1$ reach a median eccentricity of 0.54. Their orbital planes are tilted by a median of $40^\circ$ from their initial orientation, with $22\%$ becoming retrograde. In contrast, the orbital planes of the remaining 926 planets, whose eccentricities changed by less than 0.1, are tilted by a median of only $0.06^\circ$, and none become retrograde. If the host star's spin was initially aligned with the planet's orbit and has not since been reoriented \citep[for example, through star-planet tidal interactions that have been suggested to be important for hot Jupiters, but that would be negligible for cold Jupiters;][]{winn2010hot,albrecht2012obliquities,rice2022origins}, the angle between each planet's initial and final orbital planes should remain as the present-day spin-orbit obliquity of its host star. The planets excited before ionization would then appear today as eccentric, misaligned giant planets orbiting single stars, with no remaining trace of the stellar companion responsible--similar to the mechanism proposed by \citet{stephan2024novel}, where hot Jupiters are left around single stars when white-dwarf natal kicks unbind their binaries.

Planets around former wide binary components should make up only a small fraction of the planets orbiting single stars, though this fraction grows with age. Let $R(t)$ denote the ratio, at age $t$, of planets orbiting a star whose wide companion has been stripped by ionization to planets still orbiting a member of a bound wide binary. If we then assume that the field population formed at a constant rate, and then average over the duration of our simulations, we find $\langle R \rangle \approx 0.5$. Since a fraction $f_\mathrm{now} \approx 5$--$10\%$ of the planet hosts in the \citet{lammers_winn_gaia_26} catalogs are presently in wide binaries (Section \ref{sec:pop_ecc}; see also \citealt{lepine_bongiorno_2007, gaia_edr3_nearby_stars_catalog}), this implies that ${\sim}f_\mathrm{now}(1 + \langle R \rangle) \approx 7$--$15\%$ of these stars initially formed in such configurations, and that former wide binary members make up ${\sim}f_\mathrm{now}\langle R \rangle \approx 2.5$--$5\%$ of single-star giant planet hosts. This fraction depends strongly on age, however, since $R$ rises from ${\sim}0.04$ at 0.5\,Gyr to ${\sim}0.85$ at 8.5\,Gyr.

Former wide binary members should therefore make up $\lesssim 0.5\%$ of single-star giant planet hosts at ages of $\lesssim0.5$\,Gyr, but ${\sim}3$--$7\%$ in the oldest, a non-negligible contribution. Such planets may also complicate the dynamical interpretation of individual systems whose high eccentricities or large stellar obliquities are otherwise difficult to account for.

\section{Discussion} \label{sec:discussion}

In this Section, we describe some caveats, limitations, and implications of our work, and we discuss potential follow-up directions.

\subsection{Planetary Initial Conditions}

We have implicitly assumed that planets hosted by stars with wide binary companions form with similar properties to those around single stars, attributing any differences between the populations to post-formation dynamical sculpting. Based on existing observations, this appears to be a reasonable first assumption. While companions within ${\sim} 100-200$\,AU suppress planet formation \citep{wang_2014_stellar_multiplicity_1, kraus_ruinous_2016}, wider companions have shown little to no observed effect on planet occurrence \citep[e.g.,][]{hirsch_2021_planets_binaires, moe_kratter_2021}. Many wide stellar companions are far too distant to meaningfully perturb planets for most of the system lifetime. Instead, as shown in this work, planets have their orbits significantly altered only during the rare epochs when the Galactic tide and stellar encounters drive the binary to a high orbital eccentricity and trigger close approaches.

To provide intuition in the simplest scenario, we have also initialized all of our systems with only one giant planet. This is, of course, not the only (or necessarily even the most common) configuration, as evidenced by our own solar system. There is no clear theoretical consensus for the typical expected multiplicity of wide-orbiting giant planets \citep[e.g.,][]{hoffman2017stochasticity, tommy_sequential_planet_formation}. Observations indicate that multiple giant-planet systems may be common, both at the population level \citep[e.g.,][]{knutson_friends_of_hj_2014, zink_howard_23} and as suggested by specific confirmed multi-planet systems like HR 8799 \citep{marois2008hr8799}, WISPIT-2 \citep{close2025wide, lawlor2026direct}, PDS 70 \citep{haffert2019pds70}, and 14 Herculis \citep[e.g.,][]{butler2003keck, bgagliuffi_2021}.

Because we consider only one planet per system, the potential influence of planet-planet interactions is not considered in our simulations. These can produce significant population-level deviations from the distributions outlined in this work. Additional planets may enable a higher rate of planetary disruptions and ejections: both impulsive \citep[e.g.,][]{rasio1996dynamical, weidenschilling1996gravitational, chatterjee2008dynamical, juric2008eccentricity} and secular \citep[e.g.,][]{laplace1775, lagrange1778, naoz2011hot, wu2011secular} interactions can profoundly sculpt the architectures of planetary systems. A future area of interest would be to consider how the effects considered in this work influence planet-planet interactions. In particular, the presence of additional planets may enable additional dynamical instabilities through outside-in cascade mechanisms \citep[e.g.][]{zakamska2004excitation, kaib_2013, yang2025third}.

On the other hand, planet-planet interactions may shield against dynamical instability: mean-motion resonances provide stabilizing influences that allow compact giant-planet systems to remain stable where non-resonant configurations would not. For instance, \cite{fabrycky_murray_clay_hr8799} demonstrated that the super-Jupiters in the HR 8799 system are too closely packed to be stable over the system's lifetime unless they are locked in a resonant chain, and resonant chains of giant planets more broadly can remain stable over long timescales around single stars \citep[e.g.,][]{nagpal_goldberg_batygin}. \citet{roisin_binary_resonance_22} showed that the disk-driven migration responsible for resonance capture in single-star systems remains effective in wide binaries: for stellar companions at $a_\mathrm{b} \approx 10^3$ AU, their simulations found that giant planet pairs were routinely captured and remained stable in resonance. However, \cite{charalambous2025long} demonstrated that stellar encounters can in principle break long-period giant resonant chains. Whether such resonant chains can remain stable in the presence of a wide binary companion perturbed by the Galactic tide and stellar encounters is therefore an open question. The answer to this question has key implications for the interpretation of astrometrically-detected exoplanet systems that are projected for discovery through Gaia DR4 and DR5, given the degeneracy between resonant systems and single, eccentric planets \citep{yahalomi_2026_resoeccentric,
greenfield_multiple_planets_gaia}.

Finally, in this work we focused on super-Jupiters near the water-ice line. This is the regime in which giant planet occurrence is expected to peak \citep{fernandes2019turnover, fulton2021cls} and in which Gaia DR4/DR5 astrometry is projected to be most sensitive to planets \citep{lammers_winn_gaia_26}. A comparison to other planetary populations, such as the more widely-separated directly imaged planets, would be an interesting subject of future work.

\subsection{Beyond the Solar Orbit}  \label{subsec:varying_galactic_orbit}

Throughout this work, we have examined only systems that follow the Sun's present-day Galactic orbit. Even in the Solar neighborhood, however, stars occupy a wide range of Galactic orbits \citep[e.g.,][]{dehnen_binney_98, gaia_dr2_katz_mw_mapping_2018,trick_2019_disc_action_space}. Thin-disk stars alone span a broad range of radial and vertical actions, and hence of orbital eccentricities and vertical excursions, which grow systematically with stellar age \citep[e.g.,][]{wielen_1977, ting_rix_2019}. The Sun is on a somewhat dynamically cooler orbit than many of its neighbors, since its velocity relative to the local standard of rest \citep{schonrich_2010} is smaller than the typical velocity dispersion of thin-disk stars of similar age \citep[e.g.,][]{nordstrom_2004}. Thick-disk stars extend this range further, with still larger vertical excursions and more eccentric orbits \citep[e.g.][]{gilmore_reid, dr3_chemical_cartography}. Moreover, stars currently near the Sun need not have always been there: radial migration may shift a star's guiding radius by ${\sim}3$\,kpc over ${\sim}9$\,Gyr \citep[e.g.,][]{sellwood_binney_2002,Frankel_2020, Zhang2026_migration, modak_hackshaw_hawkins_2026}, so many of today's local stars likely spent much of their lives at different Galactocentric radii. 

The Galaxy itself has also evolved substantially over this time. As the disk has grown, both the strength of the Galactic tide and the number density and velocity dispersion of the stars responsible for encounters have changed, so the environment a system experienced in its past may differ significantly from the one it experiences today. However, two of these effects act in opposite senses. Stars now in the solar neighborhood were, on average, born at smaller Galactocentric radii where the Galactic tide is stronger and rate of stellar encounters is higher. At the same time, the disk has grown in mass over the past ${\sim}9$\,Gyr, so the density at any fixed radius was lower in the past. These effects may, therefore, partially offset one another such that our use of a static, present-day potential is a reasonable first-order approximation. However, the exactness of the cancellation is not clear, because the Galactic disk has likely grown ``inside-out'' \citep[e.g.,][]{frankel_2019_insideout} meaning that the inner regions of the disk were already relatively dense at early times. The Galactic tide and stellar encounters that a system experiences depend not only on its present-day orbit, but on its full history within an evolving Galaxy. Our results should therefore be viewed as representative of planet-hosting wide binaries on dynamically cool orbits like that of the Sun, within a Galactic potential like that observed in the present-day.

Farther from the Solar neighborhood, these differences become more pronounced. At smaller Galactocentric radii, the Galactic tide strengthens significantly, and at Galactocentric radii $\lesssim 3.5$\,kpc, its planar component becomes comparable to or greater than its vertical component, reversing the situation in the Solar neighborhood \citep{veras_evans_2013, stegmann_24}. Since the Galactic tide's strength scales with the local dynamical density ($\rho_0$, Equation~\ref{eq:tsec}), tight binaries that appear to be only minimally perturbed in the Solar neighborhood may be significantly affected by the Galactic tide at smaller Galactocentric radii. Stellar encounters also become more frequent at the higher stellar densities of the inner Galaxy. Together, these effects should more often drive binaries to close approaches from reduced pericenter distances. At the same time, they will ionize the binaries more quickly, such that it is unclear whether planets in the inner Galaxy will be more or less strongly affected overall by Galactic perturbations than their wider-Galactic-orbit neighbors.

This question will soon become observationally relevant. Roman's Galactic Bulge Time Domain Survey is expected to discover thousands of planets through microlensing, spanning Galactocentric radii from the Solar neighborhood to the Galactic center \citep{penny_2019}, along with potentially hundreds of thousands of transiting planets, primarily in the Galactic bulge \citep{wilson_roman_transiting}. Since microlensing is sensitive to planets at roughly 1-10 AU separations, these planets will provide a natural population to test how the Galactic environment shapes the demographic-level properties of planets in binaries.

\subsection{Tidal Dissipation} \label{subsec:tides}
Due to computational constraints, we did not account for tidal dissipation in either the stars or the planet in our simulations. The influence of planetary and stellar tides is negligible for most systems examined in this work. However, in certain contexts planetary and stellar tides are indeed relevant---specifically, when the planet's orbit is driven to very high eccentricities. 

The most important consequence of this simplification is that our simulations cannot capture hot Jupiter formation via high-eccentricity migration, a process that stellar companions are often invoked to drive \citep[e.g.][]{shariat2026}. In the presence of an inclined and/or eccentric stellar companion, a planet will experience ZLK oscillations, periodically exchanging inclination and eccentricity. During high-eccentricity epochs, tidal dissipation may act to shrink and circularize the examined Jovian planets' orbits to ultimately form hot Jupiters \citep[e.g.][]{wu_murray_2003, fabrycky_tremaine_2007, wu_murray_ramsahai,naoz_hj_binary_12, petrovich_2015, anderson_2016, vick2019chaotic, liu2025double, weldon_hj_survival}.

We cannot directly estimate how many hot Jupiters would form in our simulations. However, we leverage conservation of angular momentum during the tidal dissipation process to estimate the maximum possible hot Jupiter formation rate: a planet that circularizes at fixed angular momentum settles at $a_\mathrm{f} = 2a_\mathrm{p}(1-e_\mathrm{p})$. Specifically, we treat any planet that reaches a periapsis distance $a_\mathrm{p}(1-e_\mathrm{p}) < 0.05$\,AU, corresponding to $a_\mathrm{f} < 0.1$\,AU, as a hot Jupiter candidate. Not every candidate would become a hot Jupiter, since tides must circularize the planet before it is disrupted or driven back to lower eccentricity, so the fraction of systems that produce candidates is an upper limit on the hot Jupiter formation rate.\footnote{Because we record each planet's orbit only every 0.8\,Myr, our outputs can miss a planet's closest approaches, so some candidates may go unrecorded.}

Planets that reach such close approaches are at risk of tidal disruption, so we compute two estimates of this maximum rate. The optimistic estimate counts every candidate. The conservative estimate excludes those whose periapsis falls within the Roche limit (Equation~\ref{eq:roche}) at any point in the simulation. We keep these planets in the optimistic estimate because recent work shows that close approaches to near or within the Roche limit need not completely destroy a hot Jupiter \citep{weldon_hj_survival}.

\begin{figure}
    \centering
    \includegraphics[width=1.0\linewidth]{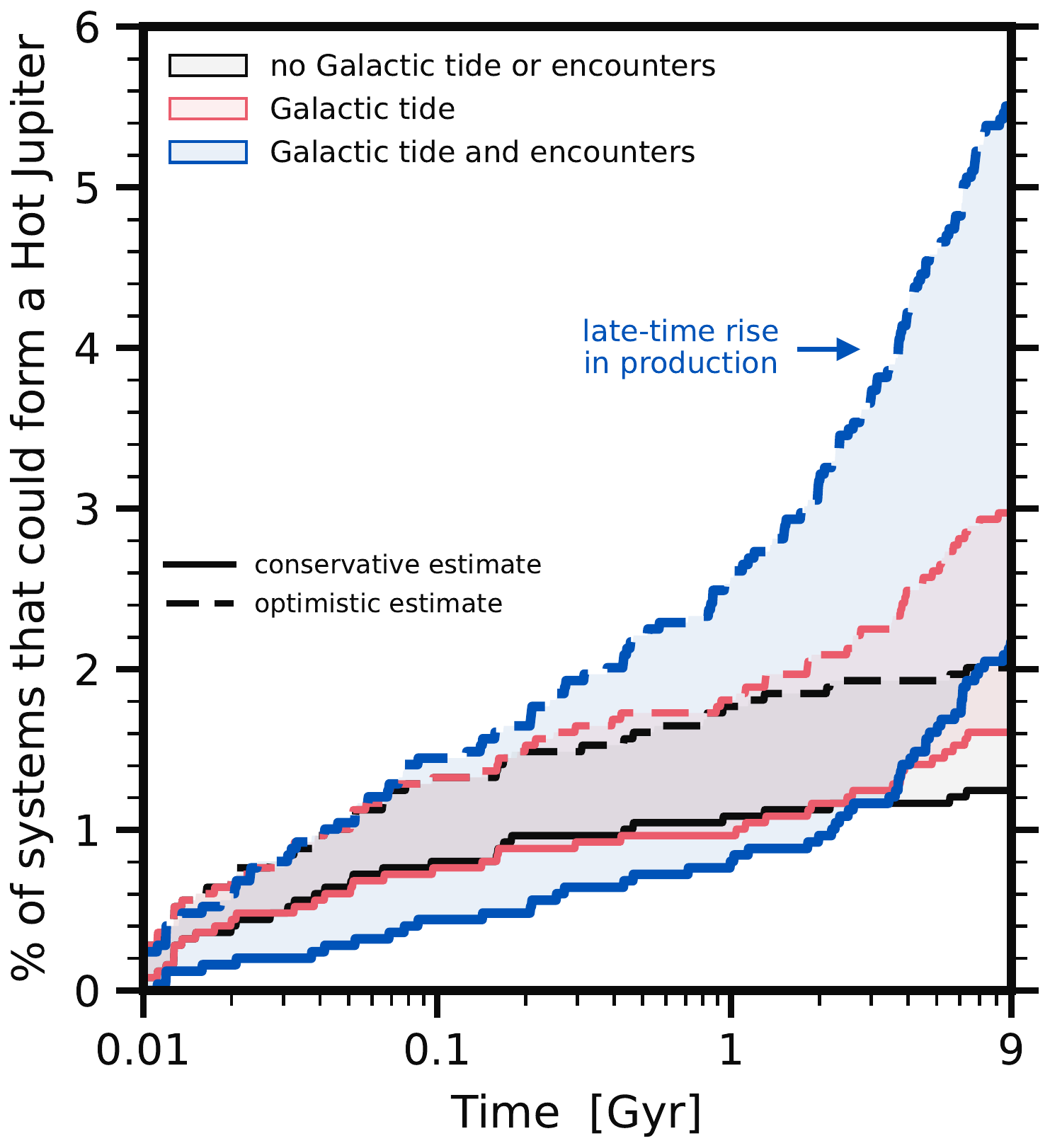}
    \caption{Upper limits on the fraction of systems in our simulations that could produce hot Jupiters, as a function of time. Black, red, and blue lines show our grids with neither the Galactic tide nor stellar encounters, with the Galactic tide only, and with both, respectively. Both curves for each grid count hot Jupiter candidates: planets that reach a periapsis small enough to tidally circularize to $a_\mathrm{f} < 0.1$\,AU (Section~\ref{subsec:tides}). Not every candidate would become a hot Jupiter, so both curves are upper limits. Solid lines (\textit{conservative estimate}) exclude candidates that are later driven within the Roche limit, assuming these planets are destroyed. Dashed lines (\textit{optimistic estimate}) include them, assuming partial tidal disruptions (which our simulations do not model) do not necessarily completely destroy a planet. By 9\,Gyr, up to ${\sim}2$--$6\%$ of systems with both the Galactic tide and stellar encounters produce candidates, many at late ages, compared with up to ${\sim}1$--$2\%$ of systems with neither effect.}
    \label{fig:hj_figure}
\end{figure}

By 9\,Gyr, the grid with both the Galactic tide and stellar encounters produces hot Jupiter candidates in ${\sim} 2\%$ (conservative) to ${\sim} 6\%$ (optimistic) of systems (Figure~\ref{fig:hj_figure}), a factor of ${\sim} 2$--$3$ more than the grid with neither effect, where candidates arise only from ZLK oscillations driven by the tightest binaries. The Galactic tide alone produces an intermediate number. This is consistent with \citet{grishin_wide_binary_hj_25}, who showed that the Galactic tide can drive hot Jupiter formation in wide binaries, and \citet{kontiainen_26} who demonstrated that stellar encounters alone may trigger high-eccentricity migration (albeit in the context of a young stellar cluster environment, rather than field stars). The maximum hot Jupiter formation rate in our grid that neglects the Galactic tide and encounters is broadly consistent with previous studies of stellar binary-driven ZLK \citep[e.g.,][]{naoz_hj_binary_12, MunozLai2016, Dawson2018}.

The Galactic tide and stellar encounters also significantly affect \textit{when} hot Jupiters form. In the absence of external perturbations, most candidates are produced within the first ${\sim} 100$\,Myr, since ZLK oscillations in the tightest binaries quickly drive susceptible planets to high eccentricity. Adding the Galactic tide and stellar encounters introduces a substantial tail of late-forming candidates (median formation time ${\sim}2.4$\,Gyr), reflecting the Gyr timescales over which these processes drive binaries to their close pericenter passages (Section~\ref{sec:planets_passages}). If even a fraction of these candidates circularize, this is one possible explanation for recent observations suggesting that a substantial fraction of hot Jupiters arrive on their orbits Gyr into their host stars' lifetimes \citep{hamer_schlaufman, chen_pnas23_hj_age, schmidt_schlaufman, chen2026origin}.

At early times, the conservative estimate for the grid with stellar encounters falls below those of the other two grids, even though the optimistic estimates are comparable. This is because encounters continually reshape the binary's orbit, so the minimum periapsis the planet reaches varies from one ZLK cycle to the next. A planet that reaches periapsis $a_\mathrm{p}(1-e_\mathrm{p}) < 0.05$\,AU early on can therefore be pushed within the Roche limit on a later cycle, which removes it from the conservative estimate. Without stellar encounters, the planet reaches nearly the same periapsis every cycle, so a planet that avoids disruption at its first close approach typically avoids it thereafter. In reality, tides would likely circularize many of these planets at their first close approach, before later cycles could disrupt them, so the conservative estimate is especially pessimistic for the grid with stellar encounters.

\section{Conclusions} \label{conclusion}

In this work, we have used $N$-body simulations to study how the Galactic tide and stellar encounters, acting through a wide binary companion, shape the long-term evolution of giant planets. We specifically examined the case of single super-Jupiters near the water-ice line (1--6.5 AU) in wide binary systems on the Sun's Galactic orbit, integrated over 9\,Gyr. Our simulations reaffirm the previously-established behavior of wide binaries under the Galactic tide, which is known to drive high-amplitude oscillations in the eccentricities and Galactic inclinations of wide binaries, as well as stellar encounters that add significant stochasticity on top of this tide-driven evolution \citep[e.g.,][]{heisler_tremaine_86, jiang_tremaine_10}. We build upon this work to demonstrate the implications of this binary evolution for companion planets.

We provide both qualitative and quantitative conclusions. The values reported are specific to our assumptions and should thus be representative of single-planet super-Jupiter systems near the ice line in wide binaries following Sun-like Galactic orbits, but may differ substantially for multi-planet systems or binaries on distinct Galactic orbits. Our simulations provide a reference point for observational comparison, and disagreements with the data may motivate the addition of further complexity to our simple baseline model. 

Our main conclusions are as follows:

\begin{enumerate}

     \item \textbf{Planets in wide binaries have their orbital eccentricities significantly excited by the Galactic tide and stellar encounters over Gyr timescales.} After 9\,Gyr, the median eccentricity of our surviving planets in wide binary systems was ${\sim}0.25$, elevated from the initial median value of ${\sim}0.15$ which we take as representative of planets orbiting single stars. The population-level eccentricity excitation grows steadily with time and is strongest in binaries with more massive companions. We therefore predict that the giant planets in wide binaries soon to be detected by Gaia DR4 and DR5 will be more eccentric than those orbiting single stars, particularly in older systems and in binaries with more massive companions.
    
     \item \textbf{Super-Jupiters near the ice line are preferentially destabilized in wide binaries with high Galactic inclinations relative to wide binaries with lower Galactic inclinations}. At high Galactic inclinations ($i_\mathrm{b} \approx 90^\circ$), the Galactic tide most efficiently drives the close passages that induce ejections and disruptions. The mutual inclinations between planetary and stellar orbits, on the other hand, remain nearly isotropic.

    \item \textbf{Planetary ejections and disruptions are common in wide binaries, with an occurrence that scales with both system age and binary mass ratio.} ${\sim}6\%$ of our planets are ejected and ${\sim}3\%$ are tidally disrupted over 9 Gyr, compared to ${\sim}1\%$ lost in total without either the Galactic tide and stellar encounters. These losses accumulate steadily with time and are most common in near-equal-mass binaries: among planets whose binary is still bound, ${\sim}20\%$ have been lost by 9\,Gyr for $q \gtrsim 0.8$, compared to ${\sim}8\%$ for $q \lesssim 0.3$. We therefore predict that giant planets should be depleted in older wide binaries and in those with more equal-mass components, including `twin' wide binaries.

    \item \textbf{Wide binaries perturbed by the Galaxy provide a steady source of free-floating planets, assuming that wide-orbiting super-Jupiters readily form in such systems.} Ejected planets leave with asymptotic speeds well described by a log-normal distribution with a median of ${\sim}4.6$\,km\,s$^{-1}$, largely retaining the Galactic kinematics of their former hosts. Ejections continue steadily over Gyr timescales, making Galaxy-perturbed wide binaries a persistent source of free-floating planets.

    \item \textbf{A small but significant fraction of present-day single-star systems with surviving planets may derive from formerly wide binary systems that were unbound by the Galactic tide and stellar encounters.} In ${\sim}43\%$ of systems, the binary is ionized while its planet survives. While this is usually caused by stellar encounters, in some cases the binary may be ionized through a close passage in which the planet itself unbinds the binary, since its orbital binding energy can exceed the binary's by more than an order of magnitude. These planets exhibit only mild orbital eccentricity excitation as a population, since the widest binaries are typically lost before they can perturb their planets. However, some planets are excited to high eccentricities and strongly tilted orbits prior to binary ionization. These systems would appear as eccentric, misaligned giant planets around single stars with no direct record of the companion that dynamically excited them.
     
    \item \textbf{Many wide binary systems become unbound by perturbations from the Galactic tide and stellar encounters over Gyr timescales.} ${\sim}22\%$ within 3\,Gyr and ${\sim}40\%$ within 9\,Gyr. We find that preferentially lower-mass binaries are ionized (${\sim}53\%$ of ${\sim}0.6\,M_\odot$ binaries versus ${\sim}30\%$ of ${\sim}2\,M_\odot$ binaries by 9\,Gyr), while the eccentricity distribution of the survivors relaxes from mildly superthermal ($\alpha \approx 1.3$) to nearly thermal ($\alpha \approx 1.0$) within ${\sim}1.5$\,Gyr, stabilizing thereafter.

    \item \textbf{The Galactic environment significantly enhances hot Jupiter formation in wide binary systems.} Between ${\sim}2\%$ and $6\%$ of systems scatter their planets onto orbits that have the potential to tidally circularize into hot Jupiters. This upper-limit hot Jupiter production rate is two to three times higher than in control simulations that exclude the Galactic tide and stellar encounters, and many hot Jupiter candidates reach the sufficiently small periapses needed for tidal circularization Gyrs after system formation. This is consistent with the findings of \citet{grishin_wide_binary_hj_25}, who explicitly modeled tidal dissipation but did not include stellar encounters, which we find further increase the production of candidate hot Jupiters.

    \item \textbf{For binaries with $a_\mathrm{b} \approx 10^3$\,AU, ZLK evolution is qualitatively altered by stellar encounters over Gyr timescales.} Binaries at these separations drive continuous ZLK oscillations in the planet's orbit. We show that for the tightest binaries in our sample, stellar encounters modulate ZLK oscillations in a manner that can more efficiently drive planets to extreme eccentricities. 
    
\end{enumerate}

Throughout this work, we have considered only systems following the Sun's Galactic orbit. Since the strength of the Galactic tide, the rate of stellar encounters, and the nature of their interplay depend on a system's orbit and location in the Galaxy \citep{veras_evans_2013, winter_petrovich_binary_mergers,stegmann_24}, extending this framework to other Galactic orbits and environments, such as the thick disk and the inner Galaxy, is a natural next step. As Gaia and Roman expand the census of planets across the Milky Way, interpreting the populations they reveal will increasingly require studying planets in the Galactic context.

\section*{Acknowledgments}

We would like to thank Alex Ji, Sarah Blunt, Suchitra Narayanan, Dan Fabrycky, Daniel Yahalomi, Thomas Wagg, Ritika Sethi, Tim Hallatt, Matthew De Furio, Quang Tran, Amir Siraj, Caleb Lammers, the Rice Research Group, the Powell Research Group, and the CCA Exoplanets Group for helpful exchanges. This work benefited from the Planets in the Galactic Context Workshop held at the Flatiron Institute and the 2026 Exoplanet Summer Program in the Other Worlds Laboratory (OWL) at the University of California, Santa Cruz, a program funded by the Heising-Simons Foundation and NASA. V.N. acknowledges support from the National Science Foundation Graduate Research Fellowship under Grant No. DGE 2140001. T.L., S.M., and A.B. are supported by Flatiron Research Fellowships at the Flatiron Institute, a division of the Simons Foundation. M.R. acknowledges support from Heising-Simons Foundation Grant \#2026-6740. This work has benefited from the use of the Bouchet computing cluster at the Yale Center for Research Computing (YCRC). This work has also used Delta at the University of Illinois Urbana-Champaign through allocation PHY260311 from the Advanced Cyberinfrastructure Coordination Ecosystem: Services \& Support (ACCESS) program, which is supported by U.S. National Science Foundation grants \#2138259, \#2138286, \#2138307, \#2137603, and \#2138296.

The generative AI tool Claude Opus 5.5 was used to suggest sentence-level wording improvements, check the manuscript for internal consistency, and to assist in writing and debugging code used for our simulations and data visualization. All suggested text and code were reviewed and verified by the authors, who take full responsibility for the content of the manuscript.

\vspace{3mm}
\noindent \textbf{Author contributions:} V.N., T.L., and M.R. jointly developed the concept and implementation of the project. V.N. led the analysis and writing for this manuscript, including all simulations. T.L. and M.R. each made significant contributions to the high-level organization and writing of the manuscript. S.M., C.H., and A.M.P-W supported this work through discussions that shaped its direction, as well as comments that refined the manuscript. D.P. and A.B. provided contributions in refining the manuscript.

\software{\texttt{gala} \citep{gala_17}, \texttt{matplotlib} \citep{hunter2007matplotlib}, \texttt{numpy} \citep{oliphant2006guide, walt2011numpy, harris2020array}, \texttt{REBOUND} \citep{rein_rebound_11}, \texttt{REBOUNDx} \citep{tamayo_reboundx_20}, \texttt{scipy} \citep{virtanen2020scipy}, \texttt{astropy} \citep{astropy_22}}

\bibliography{sample631}{}
\bibliographystyle{aasjournal}


\end{CJK*}
\end{document}